\documentclass[aps,12pt,times]{elsarticle}

\usepackage[left]{lineno}

\usepackage{amssymb}
\usepackage{amsthm}
\usepackage{lineno}
\usepackage{subfigure}
\usepackage{epsfig}
\usepackage{float}
\usepackage{pstricks}
\usepackage{tabularx}
\usepackage{supertabular}
\usepackage{ctable}
\usepackage{amsmath}
\usepackage{amssymb}
\usepackage{amsfonts}
\usepackage{stmaryrd}

\usepackage[boxed]{algorithm2e}

\usepackage{mathtools}

\usepackage[margin=0.8in]{geometry}

\newcommand {\mbs}[1]   {\boldsymbol{#1}}

\def\bD{{\mbs{D}}}    \def\bF{{\mbs{F}}}

\def\bP{{\mbs{P}}}    
    
\def\bV{{\mbs{V}}}  \def\bX{{\mbs{X}}}  
  \def\b0{{\mbs{0}}}

\def    \bD {{\mathbf{D}}}
\def    \bF {{\mathbf{F}}}

\def    \bP {{\mathbf{P}}}

\def\ba{{\mbs{a}}}    
  \def\be{{\mbs{e}}}

    \def\bu{{\mbs{u}}}
\def\bv{{\mbs{v}}}  \def\bx{{\mbs{x}}}

\def    \n1 {{{n+1}}}
\def    \Fe {{\mathbf{F}_{\rm e}}}
\def    \Fa {{{ \mathbf{F}}_{\rm a}}}

\def    \Ja {{J}_{\rm a}}

\def    \b  {{\bf b}}

\def    \n  {{\bf n}}

\newcommand{\diag}{\operatorname*{diag}}

\begin{document}

\begin{frontmatter}

\title{A modeling framework for contact, adhesion and mechano-transduction between
excitable deformable cells}

\author[a]{Pietro Lenarda}
\author[b]{Alessio Gizzi}
\author[c]{Marco Paggi}


\begin{abstract}

Cardiac myocytes are the fundamental cells composing the heart muscle. The propagation of electric signals and chemical quantities through them is responsible for their nonlinear contraction and dilatation. In this study, a theoretical model and a finite element formulation are proposed for the simulation of adhesive contact interactions between myocytes across the so-called gap junctions. A multi-field interface constitutive law is proposed for their description, integrating the adhesive and contact mechanical response with their electrophysiological behavior. From the computational point of view, the initial and boundary value problem is formulated as a structure-structure interaction problem, which leads to a straightforward implementation amenable for parallel computations. Numerical tests are conducted on different couples of myocytes, characterized by different shapes related to their stages of growth, capturing the experimental response. The proposed framework is expected to have impact on the understanding how imperfect mechano-transduction could lead to emergent pathological responses.

\end{abstract}


\cortext[cor1]{\scriptsize Corresponding author: Marco Paggi,
marco.paggi@imtlucca.it Tel: +39-0583-4326-604, Fax:
+39-0583-4326-565}
\end{frontmatter}


\section{Introduction}
\label{sec:Introduction}

Computational modeling of soft biological tissues represents a major field of
research since the last two decades, see e.g.~\cite{humphrey:2003,vairo:2013,holzapfel:2017} for fundamental textbooks and review articles.
The accurate mathematical description of excitable deformable cells and tissues has received increasing
attention due to the complexity of the involved interactions~\cite{lim:2006,lee:2007,chien:2008} in which
mechano-regulated cellular processes are key features~\cite{gao:2016,gao:2015} demanding also the development of novel experimental techniques for their characterization and understanding~\cite{ehret:2016}.
A vast literature on this topic relies on cardiovascular modeling and development~\cite{taber:2001} highlighting
as soft biological tissues, and cardiac cells in particular, span a wide range of spatial and temporal scales~\cite{qu:2014,gizzi:2016CMAME,quarteroni:2016}.

A cardiomyocyte is the functional unit of the heart at the micro-scale in which dedicated intracellular and intercellular
mechanisms allow cardiac cells to work as a syncitium~\cite{pullan:2005}.
Synchrony is the result of a fast propagation of the electrical impulse from a cell to another over the subsequent cardiac compartments. Current flow-mediated propagation
of the electrical impulse is ruled by clusters of inter-cellular proteic channels (mainly Cx43)~\cite{rohr:2004,knight:2015},
namely gap junctions (GJ)~\cite{dehin:2006,harris:2009} that are voltage- and time-dependent and can induce important effects
on the overall emerging dynamics~\cite{henriquez:2001}.
Mechanotransduction, in addition, allows cardiomyocytes to convert a mechanical signal to a change in cell growth
or remodeling~\cite{jaalouk:2009,ingber:2009}. Abnormal mechanotransduction, however, can lead to a variety of
diseases~\cite{kresh:2011,sheehy:2012} and understanding the connection between diseases and imperfect cell-cell
interactions or mechanical damage  is a cutting-edge research topic~\cite{rother:2015}.

A progress with respect to the state-of-the-art requires a novel theoretical
framework integrating modeling of biological cell interactions
within a multi-field approach~\cite{paggi:2016ICTAM}.
While this integration of methods is
still unchallenged due to the high computational and modeling
complexities, relevant contributions regard finite element procedures for the theoretical
description of single cell contractility responses under different
environmental stimuli~\cite{okada:2005,deshpande:2006,tracqui:2009,ronan:2012,ruiz:2014}
or whole reconstructed heart geometries for
selected pathological states~\cite{wong:2012}.
In order to incorporate dominant mechanisms occurring at different
scales within a constitutive framework for the cardiac tissue,
microstructural properties have to be properly described, including
mechano-regulated interactions occurring among tissue constituents.
Though our modeling refers to the cell micro-scale, we assume a
continuum approach~\cite{ruiz:2014}. Experimental
evidences on single cell contractility showed that forces are
induced where no visible stress fibers are present, thus implying
that a much finer scale is responsible for the observed phenomena
and therefore continuum level considerations can be adopted~\cite{deshpande:2006,deshpande:2008,deshpande:2015,deshpande:2016}.

\begin{figure}[htp!]
\begin{center}
	\subfigure[]{
	\includegraphics[width=0.3\textwidth]{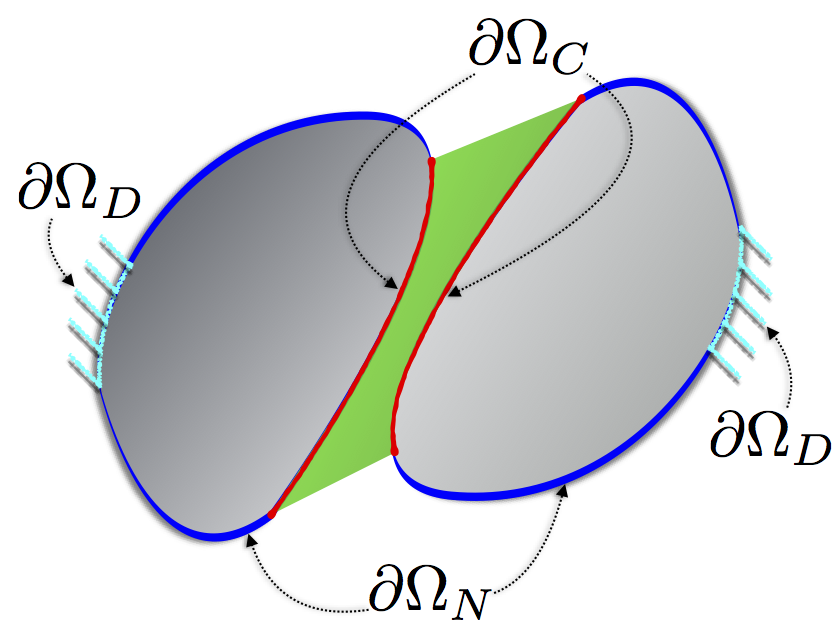}\label{fig:MDecompA}}
    \subfigure[]{\includegraphics[width=0.3\textwidth]{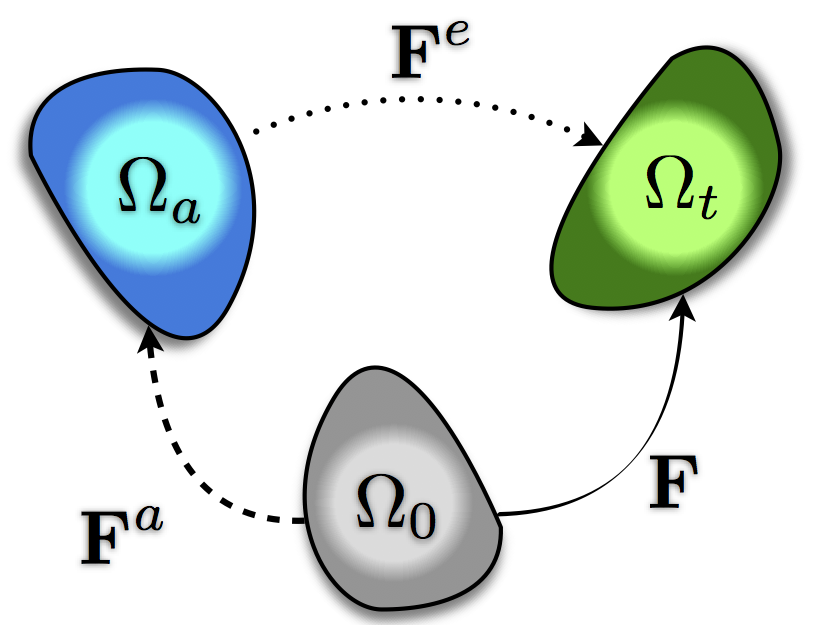}
    \label{fig:MDecompB}}
    \caption{Schematic representation of
   (a) imposed boundary conditions, e.~g., Dirichlet, $\partial\Omega_D$, Neumann, $\partial\Omega_N$, and Contact, $\partial\Omega_C$;
    (b) multiplicative decomposition of the deformation gradient in total $\bF$, active, $\Fa$, and elastic, $\Fe$, maps associated with the reference, $\Omega_0$, active, $\Omega_a$,
    and current, $\Omega_t$, configurations.
    \label{fig:MDecomp}}
\end{center}
\end{figure}

Structural
and physical properties of contact myocytes, in particular, will be
the main object of this study. Intercellular communication between excitable
contractile cells concentrates at the intercaleted discs and
concerns with microscopic electrical conductance, metabolic and
mechanical coupling~\cite{noorman:2009}. A schematic representation
of two-dimensional cardiomyocytes contact problems is provided in Fig.~\ref{fig:MDecomp}(a). The interface
constitutive model concerns (i) voltage-dependent GJs ruling the electrical conductance
for membrane voltage propagation, and (ii) adhesive and contact membrane interfaces dictating mechanical stress localization across adjacent cells.
In addition, localized focal adhesions are described via appropriate boundary conditions.

The problem at hand deserves an accurate cellular
mechanical description in which structural
heterogeneities, appropriate constitutive relations, and active
dynamics are the three key factors to be formalized within a generalized
theoretical framework~\cite{quarteroni:2016}.
We devise our myocyte contact mechanical model by linking the active
electrophysiological processes occurring in the cell with the
passive characters at the cell boundaries. The novel constitutive interface
formulation we present here does not depend on the details of the
electrophysiological and mechanical model, which we keep as simple as possible, rather we develop a
general modeling framework that can be appropriately modified and
enriched according to the selected case study.

The present work introduces two important novelties with respect to
the current literature. First, we extend the single cell study
proposed in~\cite{ruiz:2014} by formulating a novel interface constitutive model among cardiac myocytes in
electromechanical contact problems, reproducing several experimental evidences~\cite{mccain:2012}. Second, we provide a consistent
derivation of a computationally stable staggered finite element procedure for solving the two-dimensional nonlinear coupled
structure-structure electromechanical contact problem
for the interaction of two excitable deformable domains.

The paper in organized as follows.
In Section~\ref{sec:CFormulation}, the complete continuum formulation
of the active-strain model is provided.
Section~\ref{sec:CComputational} describes its weak and discretized
forms. In Section~\ref{sec:Interface}, a self-consistent theoretical
framework for the interface model is provided together with weak and
discretized forms. Section~\ref{sec:FEM}, concerns with the complete
description of the implicit scheme for finite element
implementations. In Section~\ref{sec:NumericalApplications},
numerical applications are reported and validated against
experimental evidences. Conclusions, limitations and future
perspectives are provided in Section~\ref{sec:Conclusions}.
The manuscript is equipped with extended appendices providing all the necessary derivations of the operators required for the implementation of the methodology within finite element procedures.


\section{Continuum model of the active-strain myocyte}
\label{sec:CFormulation}

In this section, the mechanical and electrophysiological model used in numerical simulations for a
single myocyte is briefly outlined. The mechanical model is based on the active-strain formulation~\cite{cherubini:2008,ambrosi:2011}.
The activation variables dynamics, which are responsible for the contraction and thickening
along the fibers vectors,  are ruled by the two-variable phenomenological
Rogers-McCulloch's model~\cite{rogers:1994}.

\subsection{Finite kinematics}

The kinematics of active deformable cells is framed within the classical
description of continuum mechanics under finite elasticity
assumptions and specialized for two-dimensional domains. In the
following, $\bX$ denotes the material position vector in the
reference (undeformed) configuration $\Omega_0 \subset \mathbb{R}^2$, and
$\bx=\bX+\bu$ stands for the spatial position
vector in the current (deformed) configuration $\Omega_t\subset
\mathbb{R}^2$ at a generic time $t$, being $\bu$ the
displacement vector. The boundary $\partial\Omega$ of the myocyte
are assumed to be Lipschitz continuous with $\partial\Omega=\partial\Omega_D\cup
\partial\Omega_N$ 
where $\partial\Omega_D$ and $\partial\Omega_N$ denote the portions of the
boundary subject to Dirichlet and Neumann boundary conditions, respectively.
The deformation gradient tensor is
$\bF=\nabla_{\bX}\bx$, and the
Jacobian of the transformation is $J=\det\bF$.

In order to encompass the nonlinear coupling between the
electrophysiological dynamics and the hyperelastic material response
induced by the excitation-contraction mechanisms in cardiomyocytes,
the well established multiplicative decomposition of the deformation
gradient into an elastic and inelastic part is assumed:
\begin{equation}\label{eq:Mdecomp}
    \bF = \Fe\Fa.
\end{equation}
Here, the active deformation gradient, $\Fa$, is provided by the
subcellular calcium/voltage dynamics. 
The sketch of the multiplicative decomposition shown in
Fig.~\ref{fig:MDecomp}(b) accounts for the intermediate non
compatible fictitious configuration $\Omega_a$ in which all the
inelastic processes take
place~\cite{cherubini:2008,gizzi:2015a}.

Let $\mathbf{a}_l$, $\mathbf{a}_t$ be the vectors of fiber sheet in longitudinal and transversal directions, respectively.
Based on the active-strain approach~\cite{nobile:2012}, the planar
active anisotropic deformation gradient tensor is built as:
\begin{equation}\label{eq:Fa}
 \Fa=\mathbf{I}+ \gamma_l \mathbf{a}_l \otimes \mathbf{a}_l + \gamma_t \mathbf{a}_t \otimes \mathbf{a}_t ,
\end{equation}
where $\gamma_l,\gamma_t$ are smooth scalar activation functions representing,
respectively, the active shortening of the cardiomyocytes in the longitudinal direction and the
corresponding thickening in the transversal direction. The activation functions are such that:
\begin{equation}\label{eq:gammat}
   \gamma_t=-\dfrac{\gamma_l}{1+\gamma_l},
\end{equation}
to achieve the
incompressibility of the cell~\cite{iribe:2007}, i.e., $\Ja=\det \Fa= (1+\gamma_l)(1+\gamma_t)=1$.

\subsection{Electrophysiological model}
\label{sec:electrophysiol}

Several electrophysiological models for the cardiac cells are available in
the literature, adopting phenomenological or
physiological approaches based on reaction-diffusion formulations~\cite{keener}.
In the present study, the Rogers-McCulloch's model
\cite{rogers:1994} is adopted for the description of the coupled model ruling the voltage potential and the Calcium concentration.
The model has been demonstrated to capture the main features of the action potential spatio-temporal dynamics with a reduced
mathematical complexity. Other possible extensions available in the literature (see e.g.~\cite{fenton:2008}) could be used for the continuum, without any loss of generality.


The nondimensional phenomenological model is
formulated as a set of nonlinear partial differential equations
describing the reaction-diffusion (RD) system ruling the coupled dynamics of dimensionless membrane voltage $V$ and the Calcium-like concentration $w$, defined in
$\Omega_0 \times [0,T]$, namely:
\begin{subequations}
 \begin{align}
\partial_t  V  + \dfrac{1}{J} \nabla \cdot \left(  \mathbf{F}^{-1} \bD  \mathbf{F}^{- \rm T} \nabla V \right)
 -  I(V,w)   & = I_{\rm app} , \\
d_t w -  H(V,w) & = 0,
 \end{align}
\end{subequations}
where $\bF^{-1}\bD \bF^{-\rm{T}}$ is the
anisotropic second order tensor of tissue conductivities obtained as the
pull-back of $\bD=\diag(D_l, D_t)$ in the planar deformed
configuration~\cite{ruiz:2014, cherubini:2008};
$\partial_t$ and $d_t$ denote partial and total derivatives in time, respectively;
$\nabla\cdot$ and $\nabla(\cdot)$ are the divergence and gradient operators, respectively;
$I_{\rm app}$ represents the external electric current.
The other electrophysiological functions are given by
\begin{subequations}
\begin{align}
I(V,w)&=c_1 V(V-a)(1-V)-c_2 V w ,\\
H(V,w)&=b(V-dw),
\end{align}
\end{subequations}
specializing the well-known FitzHugh-Nagumo model to cardiac dynamics~\cite{keener}.
Model parameters are $a=0.13$, $c_1 =0.26$ ms$^{-1}$, $c_2 =0.1$ ms$^{-1}$, $b=0.013$ ms$^{-1}$, $d=0.1$.

\subsection{Active mechanics}
\label{sec:electrophysiol}

The constitutive prescription for the free energy density reads:
\begin{eqnarray}\label{eq:energy}
    \Psi=\dfrac{\mu}{2}J_a \text{tr} \left( \Fe^{\rm T} \Fe- \mathbf{I} \right)-p(J-1),
\end{eqnarray}
where $p$ is the Lagrange multiplier arising from the imposition of the incompressibility constraint $J=1$ (conservation of mass)
and which is usually interpreted as the hydrostatic pressure field, while $\mu$ is the shear modulus.
According to Eq.~\eqref{eq:energy}, the first
Piola-Kirchhoff stress tensor reads:
\begin{eqnarray}
\label{eq:Piola}
    \bP &=&
    \mu J_a \mathbf{F} \Fa^{-1} \Fa^{- \rm T} - J p \mathbf{F}^{- \rm T}.
\end{eqnarray}
As a limit case, if coupling with the electrical variables is neglected, then the mechanical
model reduces to a standard incompressible neo-Hookean material with
strain energy density function $\Psi=\dfrac{\mu}{2}(\text{tr}(\mathbf{F}^{\rm T} \mathbf{F}) - 3)-p(J-1)$.
Again, this modeling assumption does not reduce the generality of the approach and other generalized hyperelastic models can be easily introduced.

%
%

\begin{figure}[htp]
\centering
{\includegraphics[trim=10.5cm 0cm 12.5cm 0cm, clip=true, totalheight=0.33\textwidth, angle=0]{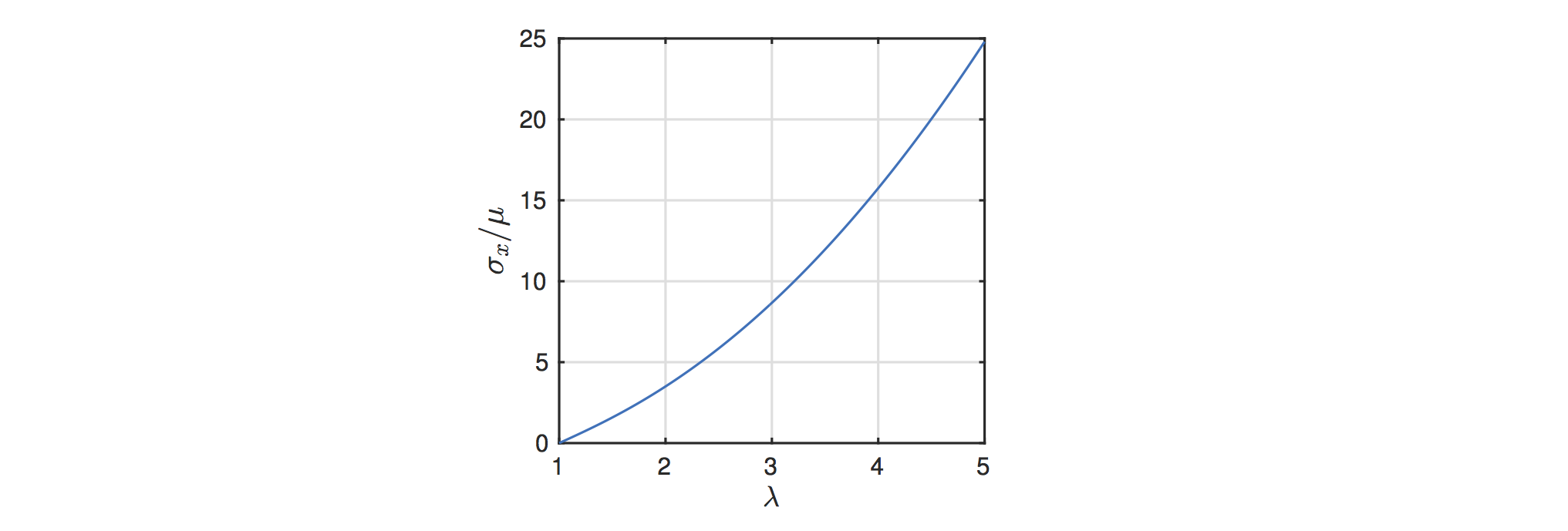}}
\caption{Dimensionless Cauchy stress component $\sigma_x/\mu$ vs. stretch $\lambda$ for different values of the shear modulus $\mu$.
\label{fig:responseS}}
\end{figure}

Let consider a unit square domain made of a neo-Hookean material, constrained on the left hand side boundary, and apply on the right hand side a constant  displacement $u_1$.
Let $\varepsilon_{x}$ be the resulting strain in the $x$-direction and
let $\lambda=\lambda_1=\varepsilon_{x}+1$ be the stretch.
The resulting Cauchy stress $\sigma_{x}$ is given by:
$$
\sigma_{x}=\mu \left( \lambda^2- \dfrac{1}{\lambda} \right).
$$
The dimensionless Cauchy stress $\sigma_{x}/\mu$ predicted
by a neo-Hookean material for various values of the shear modulus $\mu$, is shown in Fig.~\ref{fig:responseS} vs. the uniaxial stretch $\lambda$.

\subsection{Activation dynamics}
\label{sec:activation}

The mathematical formulation of the activation mechanisms in the
myofilaments adopted here is based on the arguments in \cite{cherubini:2008}. The activation dynamics is assumed to depend on
the recovery variable $w$ as:
\begin{subequations}\label{eq:activation}
\begin{align}
	f_{Ca}(w) &=
	\dfrac{1}{2} + \dfrac{1}{\pi}
	\text{atan} \left[ \beta_c \log \left( \dfrac{w}{c_0}+ \epsilon \right) \right]
	\label{activation1} ,
	\\
	\lambda_{Ca} &=
	\dfrac{f_{Ca}(c^*)-1}{f_{Ca}(c^*) - \gamma^{\max}_0}
	\label{activation2} ,
	\\
	\gamma_0 &=
	\gamma^{\max}_0 \dfrac{\lambda_{Ca}}{1+ f_{Ca}(w)}
	\label{activation3} ,
\end{align}
\end{subequations}
such that $\gamma_l=\gamma_0-1$, and $\gamma_t$ is given by Eq. \eqref{eq:gammat}. The constant $\epsilon$ is a stabilization term needed
for numerical reasons and parameters are:
$\beta_c=6$,
$c_0=3.2 \times 10^{-1} \; \rm mol/l$,
$c^*=10^{-1} \; \rm mol/l$,
$\gamma^{\max}_0=0.8$, and
$\varepsilon=10^{-5}$.

\begin{figure}[htp]
\centering
{\includegraphics[trim=6.5cm 0cm 15.5cm 0cm, clip=true, totalheight=0.33\textwidth, angle=0]{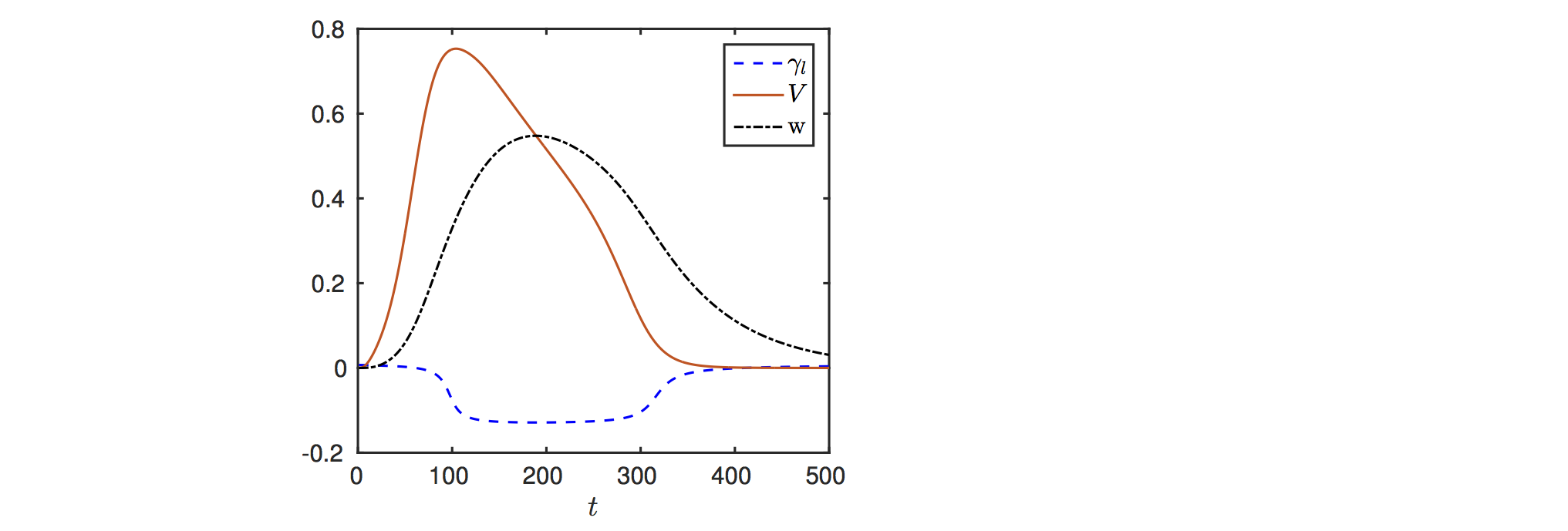}}
\caption{Time course of electrophysiological variables $V$, $w$ and $\gamma_l$ in a given sarcomere point.
}
\label{fig:response}
\end{figure}

The delay and amplitude of the Calcium-activation
interactions can be tuned based on Eq.~\eqref{eq:activation}. The described model corresponds to a
phenomenological representation of actin--myosin
binding kinetics with typical timing and Calcium function shapes
(calcium waves leading to local excitations) in agreement with more accurate models coupling
voltage and Calcium kinetics~\cite{ruiz:2014}. The reaction kinetics of Calcium
concentrations and active shortening of the sarcomeres in a
single material point behave as depicted in
Fig.~\ref{fig:response}, where a delayed activation (dashed blue line) is present according to the cytosolic Calcium concentration~\cite{rice:2008}.
On the other hand, the
present model is not capable of correctly describe
force-velocity relationships, as the microscopical
information about sarcomere dynamics is not accounted for.

\section{Computational model of  active myocytes}
\label{sec:CComputational}

\subsection{Strong and weak forms}
The strong form of the problem for a single myocyte $\Omega_0$ is given by the following set of
nonlinearly coupled partial differential equations for the displacement field $\bu$ and pressure $p$ defined in $\Omega_0$, and for the electrical
variable $V$ and the physiological variable $w$ in $\Omega_0 \times [0,T]$:
\begin{subequations}\label{strong}
\begin{align}
     - \nabla \cdot \bP & = 0, \label{strong1}  
    \\
     J &=1, \label{strong2} 
    \\
      \partial_t  V + \dfrac{1}{J} \nabla \cdot   \left(  \mathbf{F}^{-1} \mathbf{D}  \mathbf{F}^{- \rm T}  \nabla V \right)-  I(V,w)   & =  I_{\rm app} , \label{strong3} 
    \\
  d_t w -  H(V,w) &= 0. \label{strong4} 
\end{align}
\end{subequations}
The mechanical problem described by Eqs.~\eqref{strong1}\eqref{strong2} and the electrophysiological problem defined by Eqs.~\eqref{strong3}\eqref{strong4} are physically coupled
via the presence of the term $\bF_a=\bF_a(w)$ in the equation for $\bP$ accounting for the activation variables defined by Eq.~\eqref{eq:activation}, and geometrically coupled via the presence of the deformation gradient $\bF$ entering the diffusion tensor in Eq. \eqref{strong3}.
The differential problem \eqref{strong} is equipped with suitable boundary conditions describing the
interaction of the myocyte with the extracellular matrix~\cite{ruiz:2014}.
In the case of a rigid substrate, which is the case when the cell is physically isolated from the rest of the surrounding tissue~\cite{mccain:2012}, Neumann and Dirichlet boundary conditions are imposed as vanishing normal tractions on $\partial \Omega_N$ and vanishing displacements on $\partial \Omega_D$, respectively:
\begin{equation}
 \bP \, \mathbf{n}= \mathbf{0}  \,\,\, \text{on} \,\,\, \partial \Omega_N  \,, \qquad   \; \bu=\mathbf{0}  \,\,\, \text{on} \,\,\, \partial \Omega_D \, .
\end{equation}
Other boundary conditions can be considered by introducing a bed of deformable springs to simulate other stiffnesses of the extracellular matrix and support typical of experimental lob-on-chip tests~\cite{mccain:2012}.
The system of equations has to be finally completed with suitable initial values and no-flux boundary
conditions for the voltage $V$ and the electrophysiological variable $w$~\cite{cherry:2010,cherry:2011}.

Hence, in the two-dimensional setting, the formulation
contains five field variables, i.e., the displacement field
components $u_1$, $u_2$, the hydrostatic pressure $p$, the diffusive membrane voltage field $V$ and the local Calcium-like dynamics $w$.

Let now introduce the spaces where the weak solution of the problem is defined:\\
$\bu \in \bV:=L^2(0,T; [ H^1(\Omega_0)]^2)$,\;
$p \in Q:= L^2(\Omega_0)$, \;
$V \in {\mathcal V}:=L^2(0,T; H^1(\Omega_0))$, and
$w \in W:=L^2(0,T; L^2(\Omega_0))$.\\
The present formulation leads to a mixed displacement-pressure variational problem.
We also introduce the test functions for the mechanical problem
$\bv=(v_1, v_2)^{ \rm T} \in \bV_0$, $q \in Q_0$ and for the electrophysiological problem
$\xi \in {\mathcal V}_0$ and $\phi \in W_0$, defined on the spaces of the corresponding fields, and vanishing on
the Dirichlet part of the boundary. As usual, these functions multiply the strong form equations and the result is integrated over the myocyte domain $\Omega_0$.
For the sake of simplicity, we report here the weak form for homogeneous Dirichlet
boundary conditions. Applying the divergence theorem, the resulting weak problem is: Find $\bu \in \bV$, $p \in Q$ and $v \in {\mathcal V}$, $w \in W$
such that the following system is satisfied:
\begin{subequations}\label{eq:weak}
\begin{align}
&\int_{\Omega_0}  \mu J_a \mathbf{F} \Fa^{-1} \Fa^{- \rm T} : \nabla \bv \; {\rm d} \bX  -
\int_{\Omega_0}  J p \mathbf{F}^{- \rm T}  : \nabla \bv \; {\rm d} \bX   =0, \quad \forall \bv \in \bV_0,  \label{eq:Wstress}
\\
&\int_{\Omega_0} (J-1)q \; {\rm d} \bX =0 , \quad \forall q \in Q_0, \label{eq:Wpress}
\\
& \int_{\Omega_0} \partial_t  V \xi \; {\rm d} \bX+
\int_{\Omega_0} \dfrac{1}{J} \mathbf{F}^{-1} \mathbf{D} \mathbf{F}^{- \rm T} \nabla V \cdot \nabla \xi \; {\rm d} \bX  -
 \int_{\Omega_0}  I  \xi \; {\rm d} \bX  - \int_{\Omega_0}   I_{ \rm app} \xi \; {\rm d} \bX =0 , \quad \forall \xi \in {\mathcal V}_0, \label{eq:RDintv}
 \\
& \int_{\Omega_0}  d_t  w \phi \; {\rm d} \bX- \int_{\Omega_0}  H \phi \; {\rm d} \bX  =0, \quad \forall \phi \in W_0. \label{eq:RDintw}
\end{align}
\end{subequations}
The nonlinear coupled system of equations \eqref{eq:weak}, equipped with the activation equations \eqref{eq:gammat} and \eqref{eq:activation} is solved numerically using a finite element procedure, as detailed in Appendix~\ref{sec:Appendix}.
Following previous studies~\cite{quarteroni:2016}, the solution strategy
is based on a sequential (staggered) approach.
This procedure consists in dividing the problem into a mechanical phase, corresponding to the equations \eqref{eq:Wstress} and \eqref{eq:Wpress},
and an electrophysiological phase, corresponding to Eqs. \eqref{eq:RDintv} and \eqref{eq:RDintw}. At each time step $t$, the mechanical problem is solved first, and then, with the computed values of displacement and pressure, the
electrophysiological problem is solved. Both the mechanical and the electrophysiological problems are nonlinear, so that they have to be embedded into a Newton-Raphson
iterative scheme to find a solution (see Appendix~\ref{sec:Appendix} for more details).

\subsection{Numerical validation of the electromechanical model for a single myocyte}
In order to assess the accuracy of the numerical scheme described in the previous section, a series of numerical tests is performed. For the sake of simplicity, we consider equal diffusivity in the longitudinal and transverse direction, i.e. $D_l=D_t=10^{-3}\,{\rm cm}^2/{\rm ms}$, and the shear modulus is set as $\mu=4$ kPa.

\begin{figure}[htb!]
\begin{center}
  {\includegraphics[trim=6.5cm 0cm 12.5cm 0cm, clip=true, totalheight=0.23\textheight, angle=0]{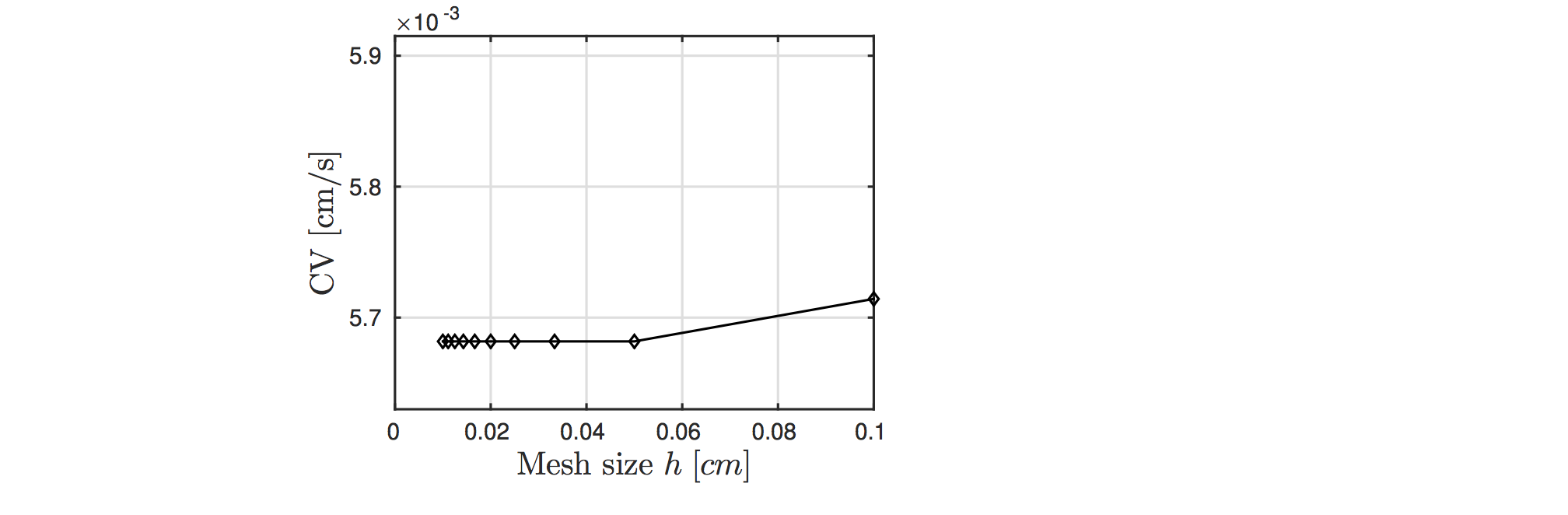}}
    \caption{Conduction velocity of the electric potential $V$ versus mesh size $h$.}
    \label{fig:CV}
\end{center}
\end{figure}

\begin{figure*}[htp!]
\centering
\subfigure[$t=49$ ms]{\includegraphics
[trim=4cm 17cm 5cm 1cm, clip=true, totalheight=0.23\textwidth, angle=0]
{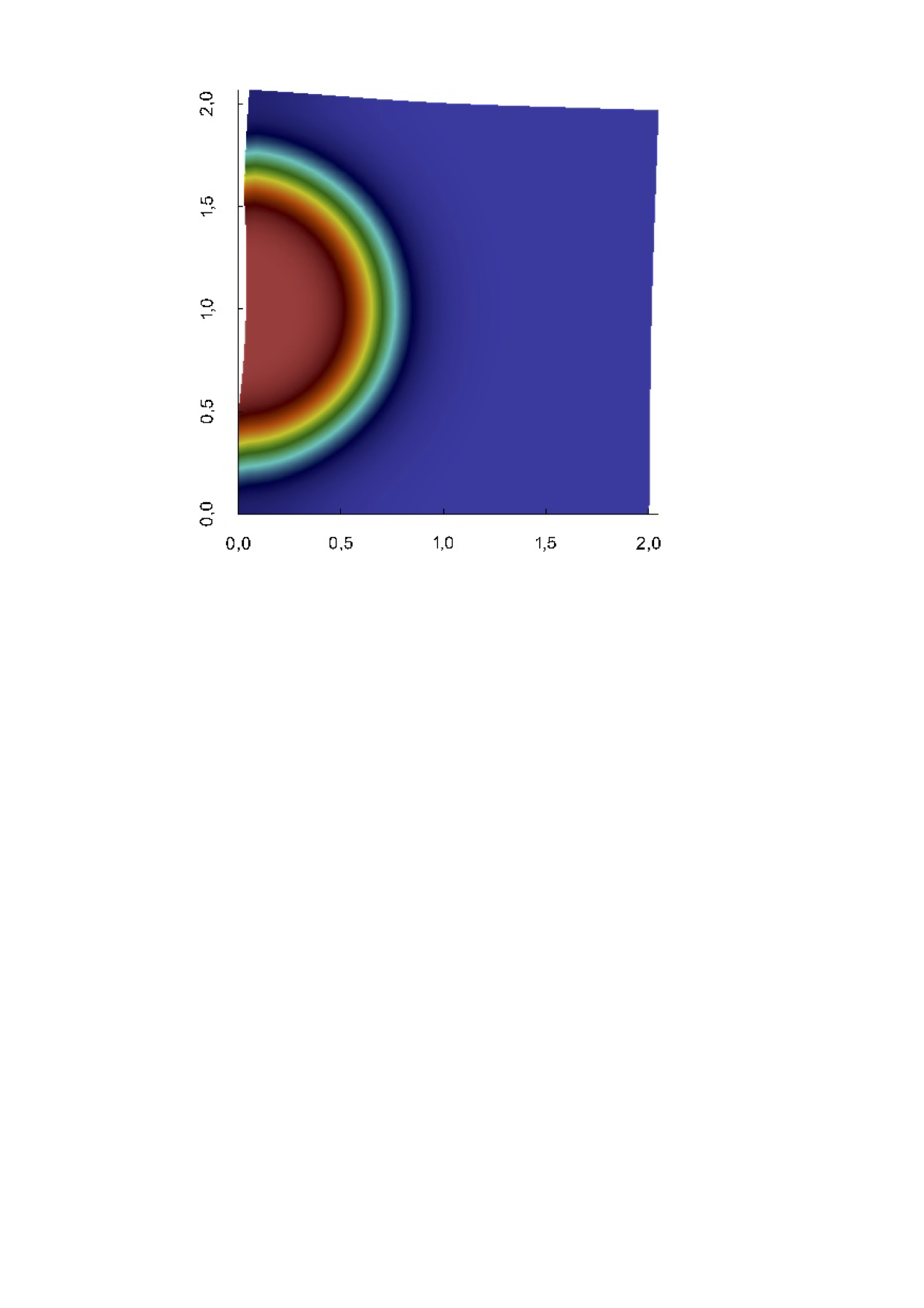}}
\subfigure[$t=135$ ms]{\includegraphics
[trim=4cm 17cm 5cm 1cm, clip=true, totalheight=0.23\textwidth, angle=0]{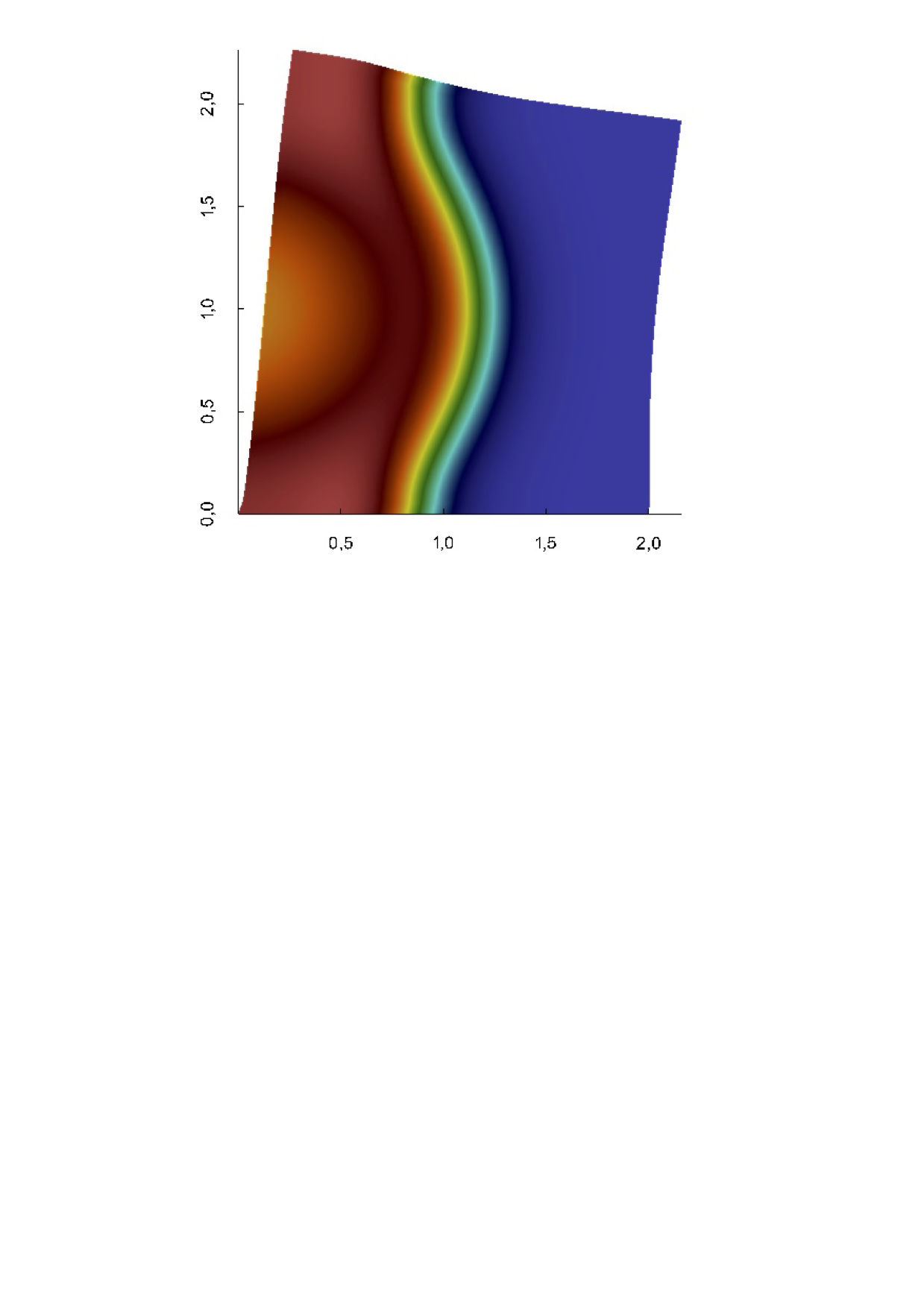}}
\subfigure[$t=250$ ms]{\includegraphics
[trim=4cm 17cm 5cm 1cm, clip=true, totalheight=0.23\textwidth, angle=0]{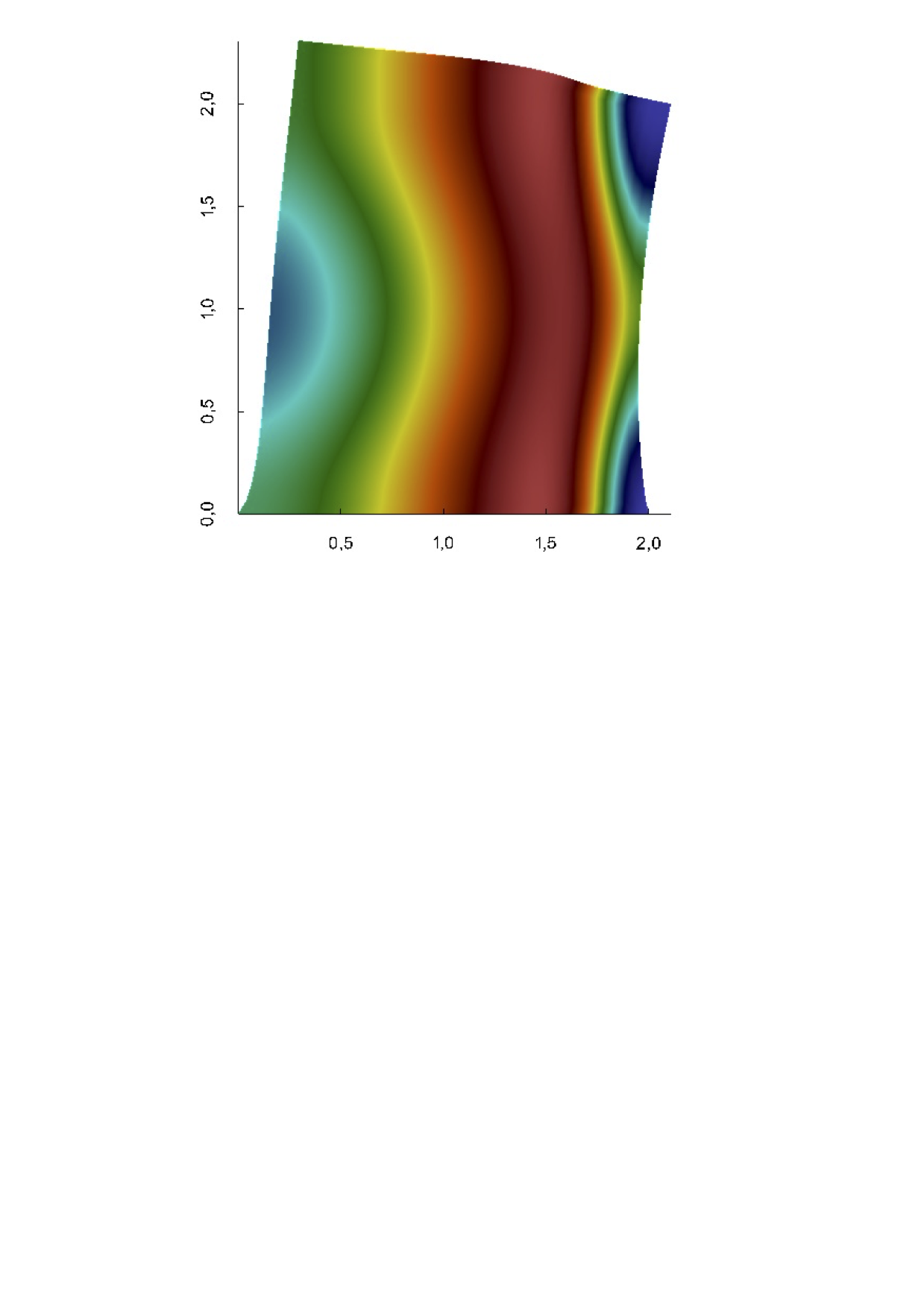}}
\subfigure[$t=350$ ms]{\includegraphics
[trim=4cm 17cm 3cm 1cm, clip=true, totalheight=0.23\textwidth, angle=0]{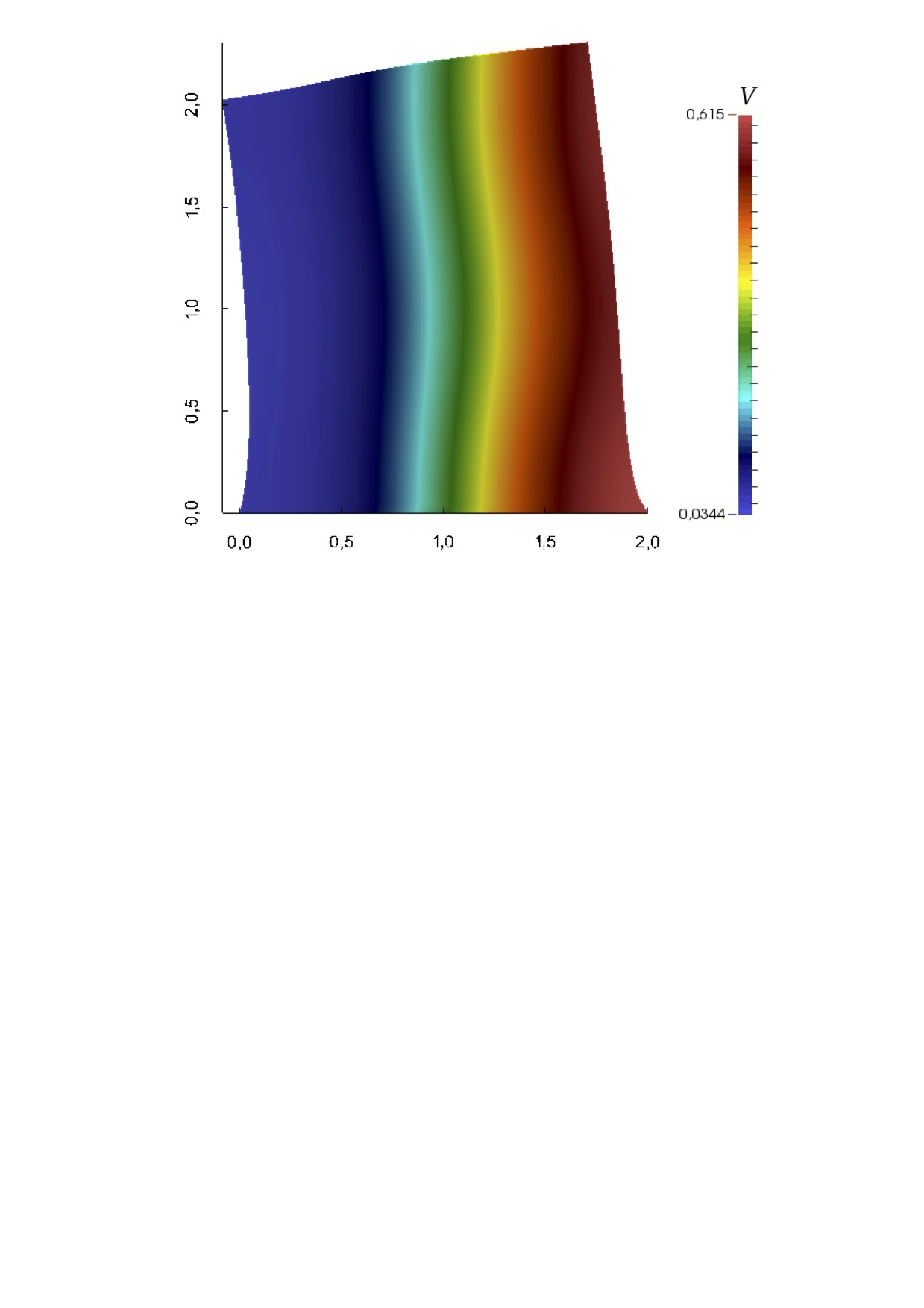}}
\caption{\label{fig:bench}
Evolution of the deformed domain and contour plots of the dimensionless action potential $V$ at different times. The color map refers to the action potential $V$.}
\end{figure*}

The first test is conducted to study the convergence of the electromechanical solver
to the physical solution as function of the mesh size $h$.
Let consider a 2D electromechanical tissue occupying in its reference configuration the squared domain $\Omega_0=[0,2]\times[0,2]$ cm$^2$.
The square is clamped on the bottom side, i.e., $\partial \Omega_D=\{ (X_1,X_2): 0 \leq X_1 \leq 2, X_2=0 \}$ and
an electrical stimulus $I_{\rm app}=\exp(-10(X^2_1+(X_2-1)^2)) $ is applied for a time interval $T_{\rm app}=[0,2]$ ms.
The fibers oriented along the Cartesian axes.
The domain $\Omega_0$ is discretized using triangular quadratic Lagrangian finite elements and homogeneous mesh size $h$. Different refinements $h \in \{ \frac{1}{10}, \frac{1}{20}, \dots, \frac{1}{100} \} \; \rm cm$ are considered.
Each simulation runs for a total time $T_{\rm fin}=500$ ms with a constant time step $\Delta t=1$ ms.
The conduction velocity (CV) of the action potential wave is
computed for different values of the mesh parameter $h$, in
order to evaluate the minimal mesh size able to accurately solve the electromechanical problem~\cite{cherubini:2012}.
Figure \ref{fig:CV} depicts the computed CV for different values of $h$, showing that the electromechanical problem results mesh-independent for $h <0.04$ cm.
The number of iterations for the mechanical and the electrophysiological problems, each one solved with a Newton-Raphson method and considering tolerances for the residual error norms of $1\times10^{-14}$, ranges between 4 and 8 during the simulation.
The evolution of the deformed domain for the test problem, with a superimposed contour plot of the electric potential $V$ at different times, is shown in Fig.~\ref{fig:bench}. The action potential wave, generated via the localized circular tissue excitation $I_{app}$,
propagates form left to right and induces non-trivial deformations of the tissue.

\begin{figure}[h!]
\begin{center}
  \subfigure[Deformed mesh at $t=200$ ms]{
  \includegraphics
[trim=4cm 16cm 2cm 2cm, clip=true, totalheight=0.3\textwidth, angle=0]{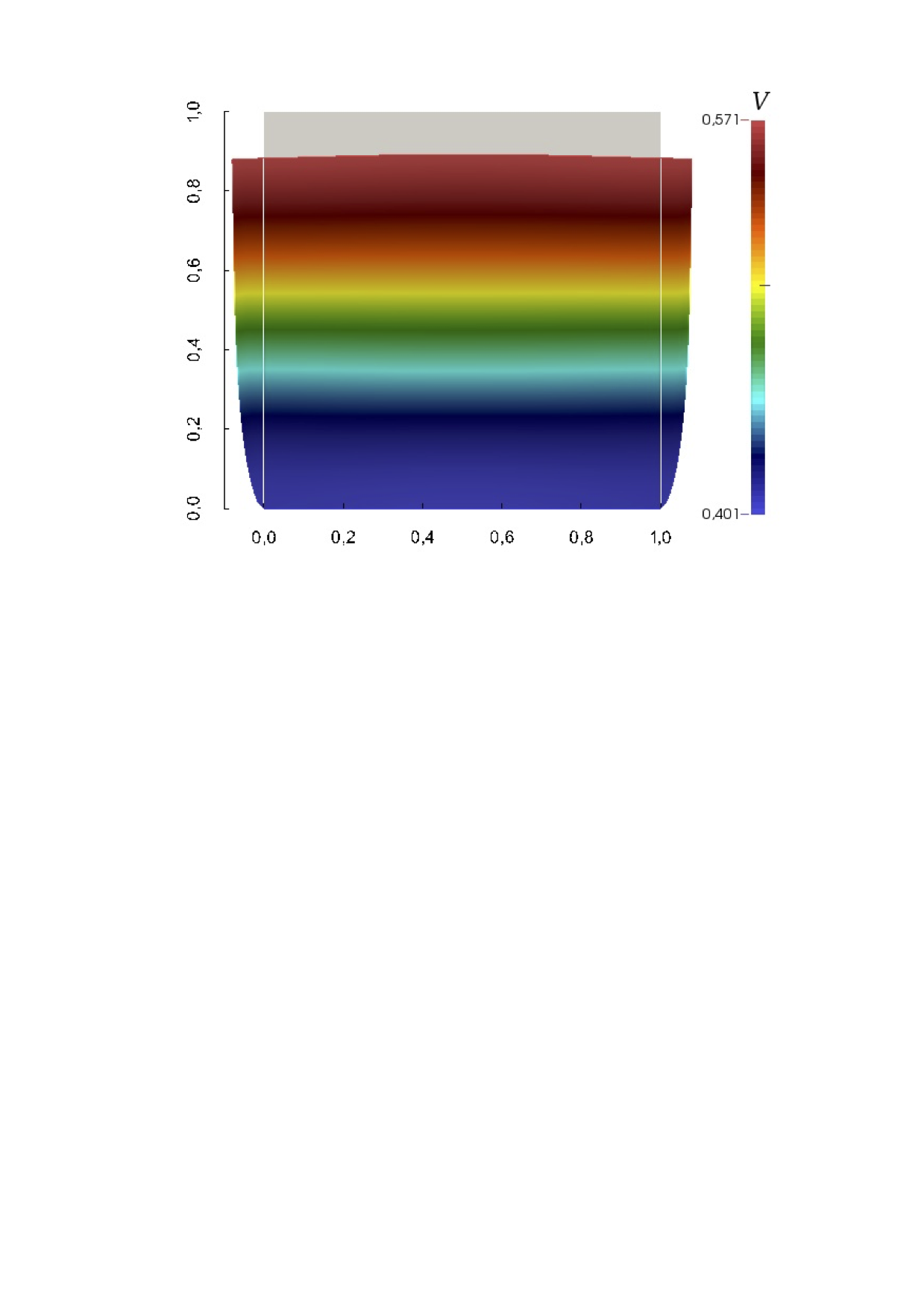}}
  \subfigure[Time evolution of the electrophysiological variables]{\includegraphics[trim=9cm 0cm 14cm 0cm, clip=true, totalheight=0.3\textwidth, angle=0]{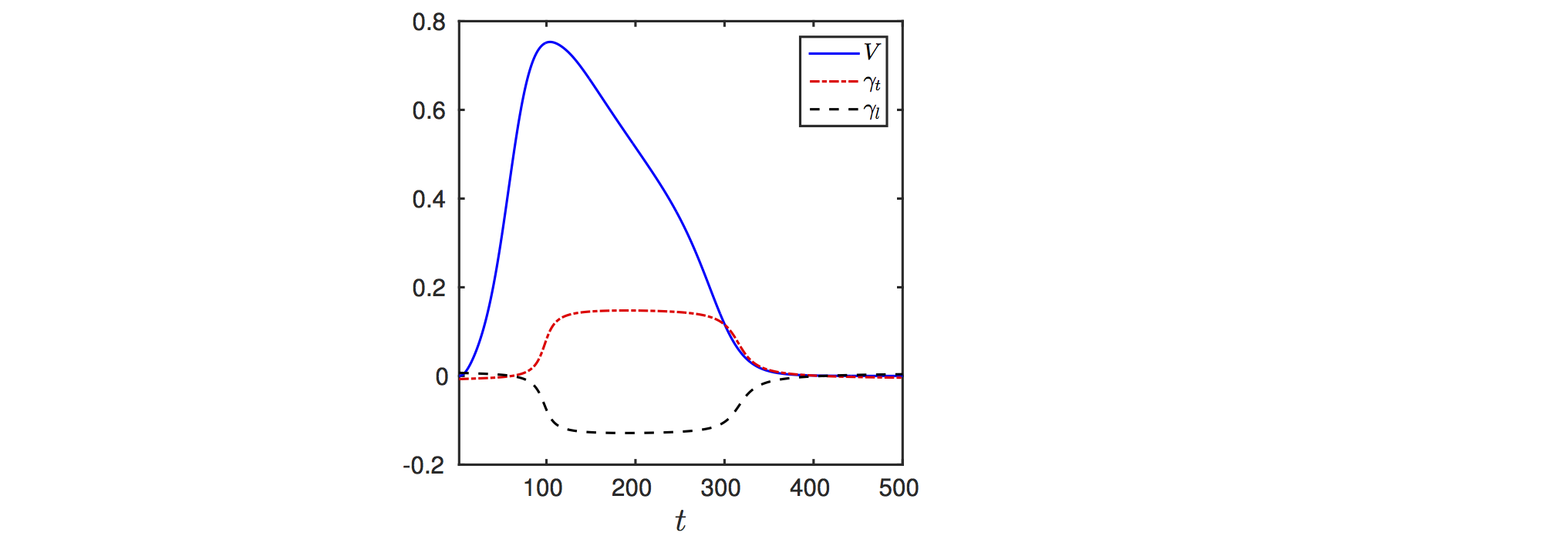} }
    \caption{ \label{fig:contraction}
    (a) Undeformed (gray) and deformed (colored) domain at $t=200$ ms, with superimposed contour plot of the electric potential $V$.
    (b) Time evolution of the dimensionless electric potential $V$ and of the activation variables $\gamma_l$ and $\gamma_t$.
}
\end{center}
\end{figure}

A second benchmark test, highlighting the role of the anisotropy induced by the activation dynamics, is performed on a square domain $\Omega_0=[0,1]\times[0,1]$ cm$^2$. The Dirichlet part of the boundary is now represented by the bottom side of the domain, i.e.,
$\partial \Omega_D=\{ (X_1,X_2): 0 \leq  X_1 \leq 1 ,  X_2 =0 \}$.
The fiber vectors, defining the underlying cell microstructure, are in this case $\ba_l=(0, 1)^{\rm T}$, $\ba_t=(-1, 0)^{ \rm T}$.
The initial values for dimensionless electric potential and Calcium concentration are null everywhere but set in the active state
$V_0=0.7$ and $w_0=0.2$ on the domain with $X_1 \leq 0.2$ cm.
These configuration is able to generate an action potential wave propagating from the bottom to the top of the domain.
As expected~\cite{nobile:2012}, the
electrical excitation wave induces a non-trivial contraction of the tissue due to the nonlinear coupling and the tissue anisotropy.
According to experimental evidence, the maximum contraction of $20\%$ of the resting length is recovered in the longitudinal direction. The deformed configuration at the peak of the activation variable $\gamma_l$ is shown in Fig.~\ref{fig:contraction}(a) with the superimposed contour plot of the electric potential $V$. Due to material incompressibility, thickening in the transverse direction is also observed.
The corresponding evolution of the action potential wave (solid line) and of the activation variables (dashed lines) from the center of the domain is shown in
Figure \ref{fig:contraction}(b).

\section{Formulation of cardiomyocytes electromechanical interaction  model}
\label{sec:Interface}

\subsection{Strong form of the electromechanical contact problem}

Let the myocyte $\Omega_1$ be surrounded by a continuum representing
the extracellular matrix or by a second myocyte $\Omega_2$, see Fig.~\ref{fig:MDecomp}(a). In this framework, the single myocyte transfers tractions, electric
signals and kinetic variables across its \emph{active} interface boundary
$\partial \Omega^1_C=\partial \Omega^2_C =\partial \Omega_C $, as
experimentally observed in~\cite{mccain:2012}.
The proposed strategy to simulate this problem is to solve the electromechanical problem for myocyte 1 and then to transfer the quantities at the interface to the other myocyte during the same time step.
This approach, which is a strategy typically used for fluid-structure interaction problems~\cite{quarteroni:2000} and that is herein adopted for myocyte-myocyte contact, is prepared over the interface elements or contact elements for its easy parallelization. This will ensure the scalability of the approach for the study of contact simulations with multiple myocytes, which is the target of interest for understanding the emergence of collective properties~\cite{mertz:2012} reflected into cardiac diseases.

In order to stimulate the myocyte-myocyte interaction, an external excitation current $I_{\rm app}$ is applied to myocyte $\Omega_1$ such that a propagating front passes from $\Omega_1$ to $\Omega_2$ via $\partial \Omega_C$ (or viceversa by exciting $\Omega_2$ first, without any loss of generality).
The objective consists in finding displacements, pressures and electrophysiological quantities for $\Omega_1$ and $\Omega_2$ which are solution of
the following system of differential equations and incompressibility conditions ($k=1,2$):
\begin{subequations}
\begin{align}\label{strongo1}
    - \nabla \cdot \bP(\bu_k) &= \mathbf{0},
    \\
    J_k &=1 ,
    \\
     \partial_t   V_k + \dfrac{1}{J_k} \nabla \cdot   \left(  \mathbf{F}^{-1}_k \mathbf{D}  \mathbf{F}^{- \rm T}_k  \nabla V_k
     \right)-  I(V_k,w_k)   &=  I_{ \rm app},
    \\
    d_t w_k -  H(V_k,w_k) &= 0,
\end{align}
\end{subequations}
where $\bu_k, p_k$ are the displacement and pressure fields defined in $\Omega_k$ while $v_k, w_k$ are the electric potential and Calcium concentration
defined in $\Omega_k \times [0,T]$.

\begin{figure}[ht!]
\centering
{\includegraphics[trim=1.8cm 11cm 0cm 9cm, clip=true, totalheight=0.3\textwidth, angle=0]{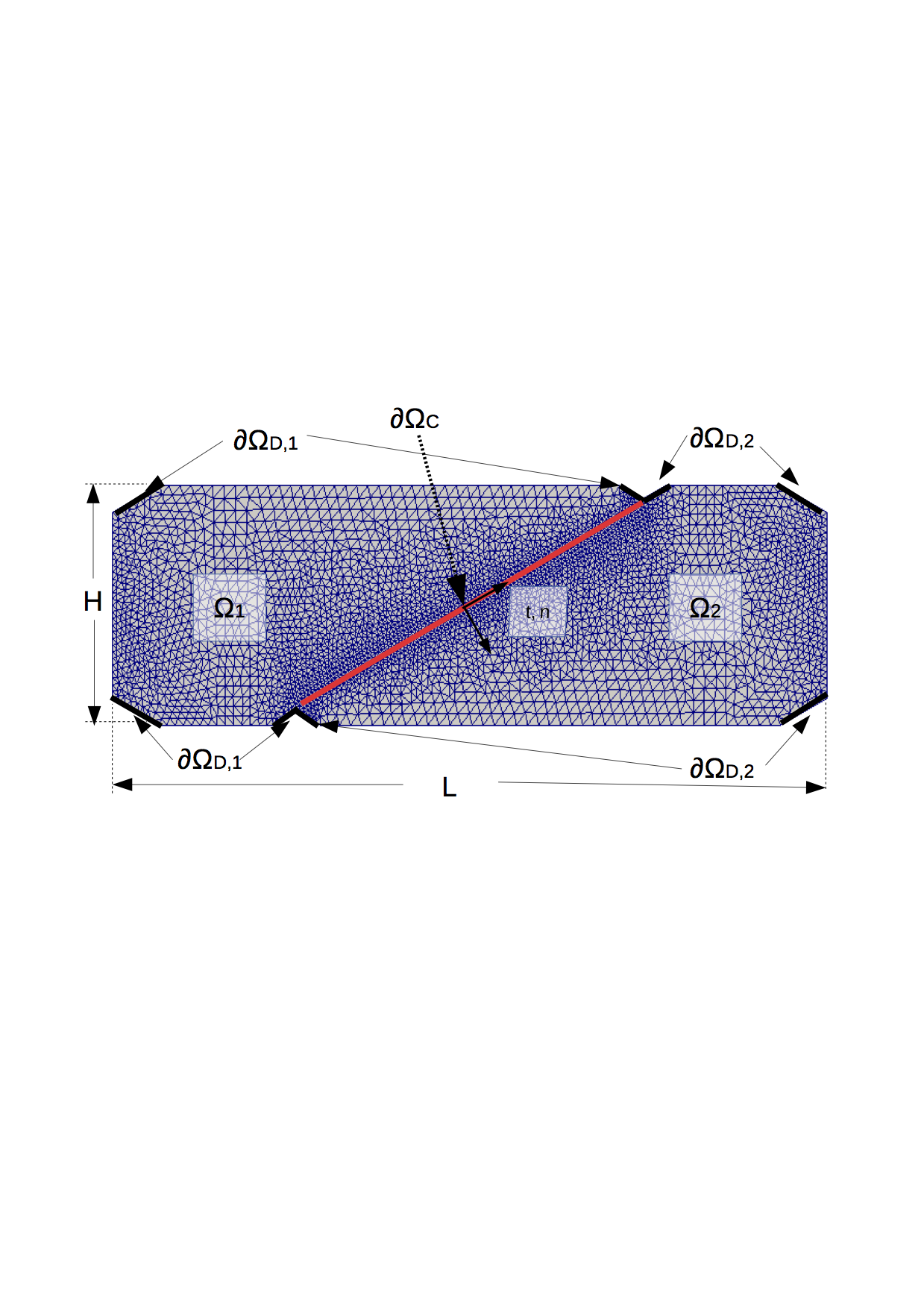}}
\caption{Sketch of the mesh and the Dirichlet boundary conditions used in the numerical simulation for
the electromechanical contact problem between two cardiomyocytes.}
\label{fig:mesh}
\end{figure}

Homogeneous displacement boundary conditions are prescribed on the Dirichlet portions of the domains $\Omega_{D,1}$ and $\Omega_{D,2}$, see Fig.~\ref{fig:mesh}:
\begin{equation}\nonumber
 \bu_1=\mathbf{0}  \,\,\, \text{on} \,\,\, \partial \Omega_{D,1} \quad \text{and} \quad
 \bu_2=\mathbf{0}  \,\,\, \text{on} \,\,\, \partial \Omega_{D,2} \,.
\end{equation}
In addition, we assume the two electromechanical domains exchange tractions and electrical signals through the common interface boundary $\partial \Omega_C$, which can be mathematical formulated as mixed boundary conditions:
\begin{equation}\label{eq:BC}
\begin{cases}
  \mathbf{P}\,\mathbf{n}_{12} + \mathbf{T}_{12}  &= \mathbf{0} \\
  \mathbf{P}\,\mathbf{n}_{21} + \mathbf{T}_{21}  & = \mathbf{0}
 \end{cases}  \quad \text{on} \quad  \partial \Omega_C,
\end{equation}
\begin{equation}\label{eq:BC1}
\begin{cases}
 \nabla \cdot \left( \mathbf{D} \nabla  V_1 \right) \cdot \mathbf{n}_{12} + c D_n (V_1-V_2)=0 \\
  \nabla \cdot \left( \mathbf{D} \nabla V_2 \right) \cdot \mathbf{n}_{21} + c D_n (V_2-V_1)=0
 \end{cases}
\ \text{on} \quad  \partial \Omega_C,
\end{equation}
implicitly assuming Neumann zero flux conditions for the Calcium concentration field $w$ according to the formulation proposed in the previous section.
Here $\mathbf{T}=\mathbf{T}_{12}=-\mathbf{T}_{21}$ is the interface traction;
$\mathbf{n}=\mathbf{n}_{12}=-\mathbf{n}_{21}$ and $\mathbf{t}=\mathbf{t}_{12}=-\mathbf{t}_{21}$ are the normal and tangential unit vectors at
each point of the internal boundary $\partial \Omega_C$;
$\Delta V= \vert V_1-V_2 \vert$ is the voltage gap (transjunctional voltage) between $\Omega_1$ and $\Omega_2$; $c$ represents a so far corrective term introduced in the computational model and $D_n$ is the nonlinear contact conductance detailed in the next section.

The generalized boundary conditions \eqref{eq:BC} and \eqref{eq:BC1} complete the structure-structure electromechanical interaction problem between two adjacent myocytes $\Omega_1$, $\Omega_2$ and tacking place at the interface boundary $\partial\Omega_C$.

\subsection{Nonlinear constitutive mechanical contact model}

As far as the mechanical response of the interface is concerned, tractions exchanged
at the interface must be continuous for equilibrium considerations
\cite{PWJMPS:2012} and are assumed to be functions of the relative displacements
between the two myocytes. Experimental
results based on atomic force microscopy~\cite{benoit:2002,friedrichs:2010} clearly
highlight the existence of a contact regime in compression, and of an adhesive response in traction which can be modeled as
a nonlinear traction-separation constitutive relation in accordance with~\cite{paggi:2006,paggi:2011,paggi:2015,paggi:2016}.
To distinguish between the normal and tangential response at the interface,
the relative displacements $\bu_1- \bu_2$ are
projected along $\mathbf{n}$ and $\mathbf{t}$, providing the normal
and the tangential relative displacements defined as $g_n=(\bu_1-\bu_2) \cdot \mathbf{n}$ and
$g_t=(\bu_1-\bu_2) \cdot \mathbf{t}$,
respectively.
Accordingly, interface tractions are decomposed as $\mathbf{T}=T_n \mathbf{n}+ T_t \mathbf{t}$.

The normal component of the interface traction is a function of the interface opening $g_n$, considering a penalty model in compression ($g_n<0$) and a linear tension cut-off adhesive response in tension ($0<g_n<g_{n,{\rm max}}$):
\begin{equation}\label{eq:Tn}
T_n=
\begin{cases}
\alpha g_n , \ & \ \text{if} \quad g_{n}  < 0,
\\
T_{n, \rm max} \left( \dfrac{g_n }{g_{n ,\rm max}} \right) , \ & \ \text{if} \quad 0<g_n  <g_{n ,\rm max}
\\
0 , \ & \ \text{if} \quad g_{n}  \geq g_{n ,\rm max} ,
 \end{cases}
\end{equation}
where $g_{n, \rm max}$ is the critical separation and $T_{n,\rm max}$ is the peak adhesive strength based on experimental observations~\cite{benoit:2002}.
We can select $g_{n, \rm max}=1$ $\mu$m, $T_{n,{\rm max}}= 1000$ kPa and $\alpha=1000$ kPa for myocytes.

\begin{figure}[ht!]
\begin{center}
  \subfigure[]{\includegraphics
  [trim=7.5cm 0.7cm 15.5cm 1cm, clip=true, totalheight=0.3\textwidth, angle=0]
  {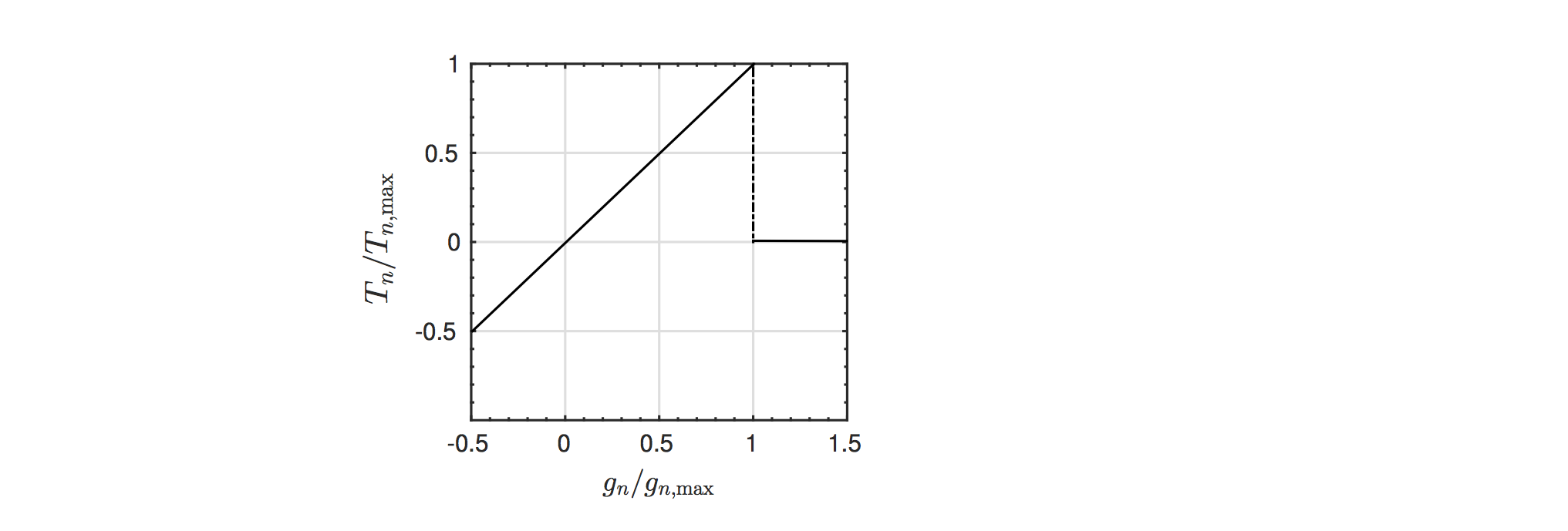}}
  \subfigure[]{\includegraphics
  [trim=7.5cm 0.7cm 15.5cm 1cm, clip=true, totalheight=0.3\textwidth, angle=0]
  {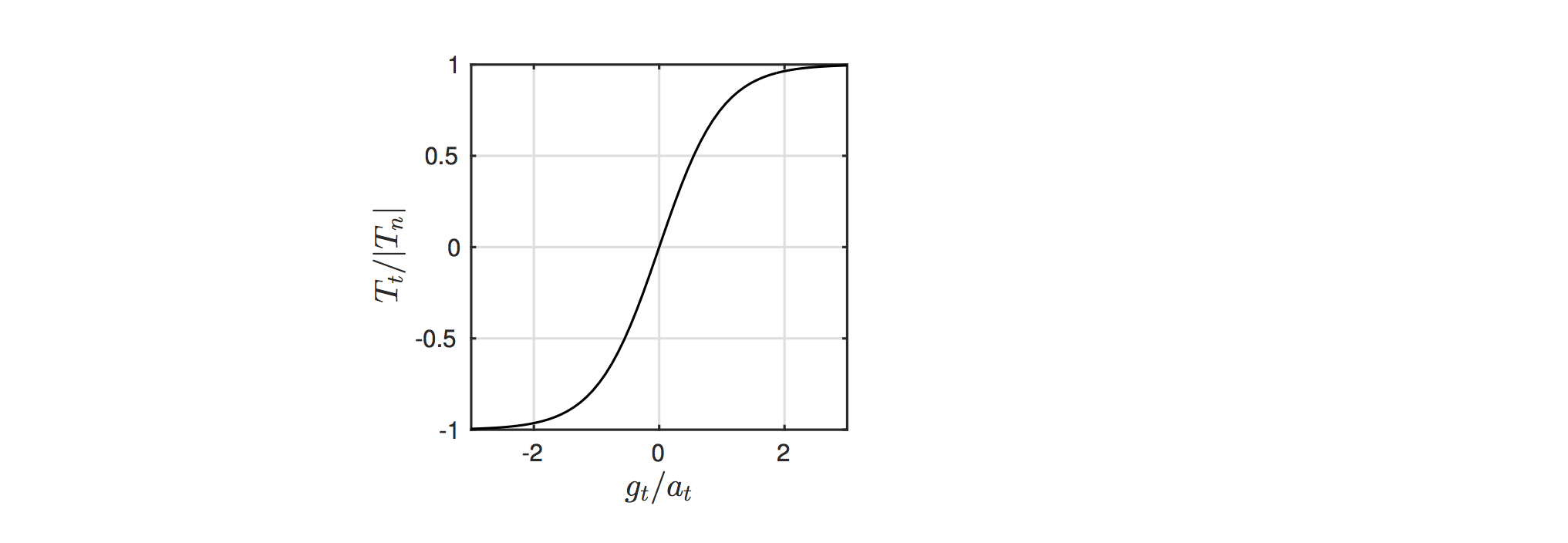}}
    \caption{
    (a) Cohesive traction-separation law.
    (b) Friction law.
    }
    \label{fig:DDV}
\end{center}
\end{figure}

The constitutive response in tangential direction is modeled as a regularized Coulomb-friction law~\cite{Wriggers:2006} in compression:
\begin{equation}\label{eq:Tt}
 T_t=f \tanh \left( \dfrac{g_{t} }{a_t} \right) \vert T_{n} \vert \,,
\end{equation}
where $f$ denotes the local friction coefficient and $a_t$
is the regularization length scale~\cite{Wriggers:2006}.
For the present problem, we set $\mu=4$ kPa, $f=1$ and $a_t=1\,\mu{\rm m}$.
In tension, tangential tractions are assumed to be negligible.
Tractions $T_n$ and $T_t$ are shown in
Figs.~\ref{fig:DDV}(a) and Fig.~\ref{fig:DDV}(b), respectively.

\subsection{Nonlinear electric conductance}

\begin{figure}[ht!]
\begin{center}
	\includegraphics
	[trim=8.5cm 0.7cm 14.5cm 1cm, clip=true, totalheight=0.3\textwidth, angle=0]{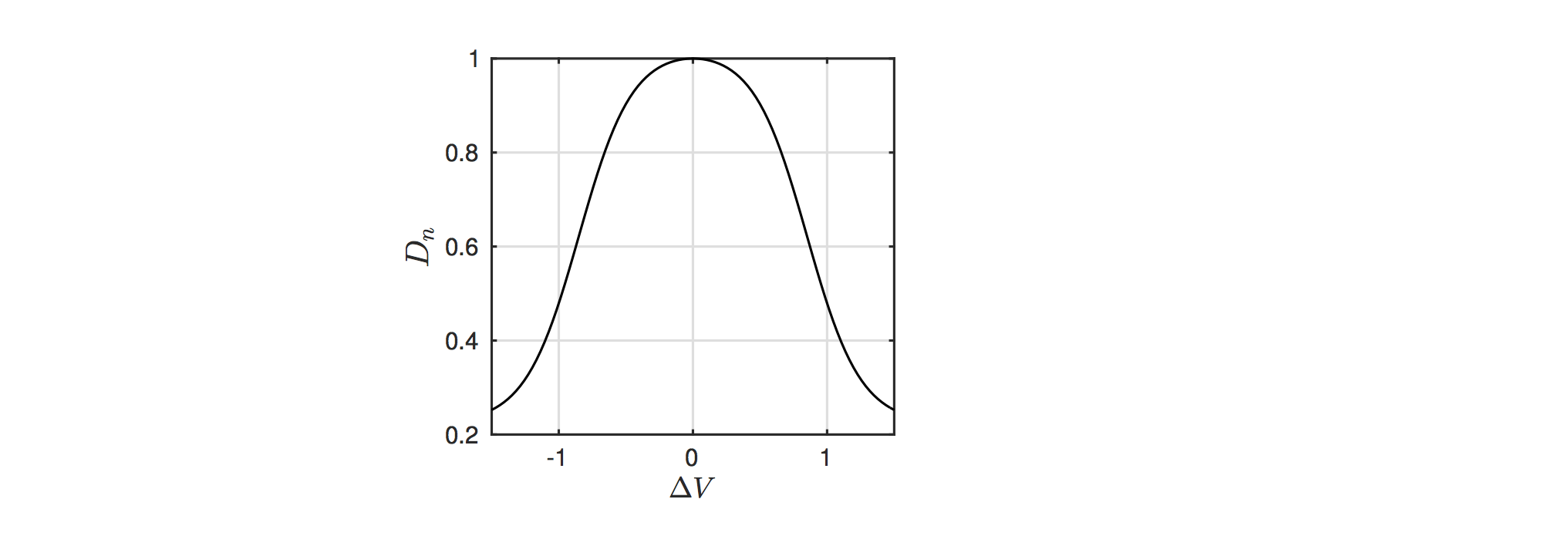}
    \caption{Conductivity per unit interface length vs. the nondimensional transjunctional voltage $\Delta V$.}
    \label{fig:Dn}
\end{center}
\end{figure}

As far as the transfer of electric voltage and current across the interface $\partial \Omega_C$ is concerned, it is assumed that the current flows solely in the direction normal to the interface.
According to the experimental evidence~\cite{dehin:2006}, the contact flux is a nonlinear function of the transjunctional voltage gap in compression and it is vanishing in tension (see also \cite{sapora:2014} for similar modeling in the case of thermo-elasticity problems at interfaces).
The phenomenological constitutive law for $D_{n}$ during contact is assumed to mimic the average steady-state
conductivity measured in dual patch clamping experiments for cardiovascular cells~\cite{chen-izu:2001}.
This quantity has the dimensions of a conductance per unit interface length,
i.e. $\mu$S/$\mu$m:
\begin{equation}\label{eq:Dn}
  D_n=a_1+a_2  \left( \dfrac{1}{1+e^{-a_3-a_4 \Delta V}} -\dfrac{1}{1+ e^{-a_3+a_4 \Delta V }} \right) \,,
\end{equation}
where
$a_1=0.2225$ $\mu$S/$\mu$m,
$a_2=0.8$ $\mu$S/$\mu$m,
$a_3=5$,
$a_4=4.25$.
The selected parameters induce 80\% of variation in the transjunctional conductance with respect to a baseline value, $a_1$
according to~\cite{chen-izu:2001}.
The constitutive prescription \eqref{eq:Dn} is depicted in Fig.~\ref{fig:Dn}.

\section{Computational modeling of the structure-structure electromechanical interaction}
\label{sec:FEM}
The numerical algorithm for the structure-structure interaction problem between two cardiac myocytes is herein detailed and implemented via the open source software FreeFem++ \cite{freefem}.
Typically, contact problems like those in the present study could be solved using node-to-segment or segment-to-segment contact formulations~\cite{Wriggers:2006}. Alternatively, since the pairing between the slave and the master segments does not change during the deformation process, interface elements could be used as well, as in~\cite{paggi:2016}.

However, in order to develop a solution scheme of easy parallelization which allows for large scale computations with several interacting myocytes, in this study we formulate the contact problem as a structure-structure interaction model. In this context, by exploiting a sequential (staggered) solution algebraic scheme, 4 nonlinear algebric systems,
resulting from the introduction of the FE discretization have to be solved: one mechanical and one electrophysiological for
myocytes 1 and one mechanical and one electrophysiological for
myocytes 2.
Coupling between the two myocytes is achived via the structure-structure boundary conditions in Eqs.~\eqref{eq:BC} and \eqref{eq:BC1}.
At each time step $t^{n+1}$, the following operations are made:
\begin{enumerate}
 \item (M1) Solution of the mechanical problem for the myocyte $1$. The superscript
 $( \cdot)^{n+1}_k$ means evaluation in the current Newton-Raphson approximation.
{The output variables are the displacement field vector $\bu^{n+1}_1$ and the
hydrostatic pressure $p^{n+1}_1$, which is a Lagrange multiplier arising from the imposition of the incompressibility constraint $J_1=1$.}
Let $\gamma^{n+1}_{l,1}$ and
$\gamma^{n+1}_{t,1}$ be the activation variables computed according to Eq.~\eqref{eq:activation} and \eqref{eq:gammat}. Given the values of
the displacement field vector and pressure $(\bu^{n}_1, p^n_1)$, and the same quantities at the $k$-th iteration of the
Newton-Raphson scheme $(\bu^{n+1}_{1,k}, p^{n+1}_{1,k})$ for the myocyte 1, and
the value of the displacement $\bu^n_2$ for myocyte 2 at the previous converged iteration, the current
normal and tangential gaps, $g^{n+1}_{n,k}=(\bu^{n+1}_{1,k}- \bu^{n}_2) \cdot \mathbf{n}$
and $g^{n+1}_{t,k}=(\bu^{n+1}_{1,k}- \bu^{n}_2) \cdot \mathbf{t}$,
and the cohesive traction vector $\mathbf{T}^{n+1}_k$ are computed. At
each iteration $k+1$, the mechanical problem of contact between the two myocytes
$1$ and $2$ consists in
finding the correction in the displacement vector and
pressure $(\delta \bu_1, \delta p_1)$ inside
myocyte $\Omega_1$ such that for all
test functions $(\bv_1, q_1)$ are solutions of the following linearized system of equations,
equipped with boundary conditions accounting
for the stucture-structure interaction between myocytes $\Omega_1$ and
$\Omega_2$ (Eqs. \eqref{eq:Tn} and \eqref{eq:Tt}) on the internal boundary $\partial \Omega_C$:

\begin{equation}\label{eq:myo1}
\begin{aligned}
&\mathcal{M}_1(\bu^{n+1}_{1,k}, p^{n+1}_{1,k}; \bu^{n}_{2})(\delta \bu_1, \delta p_1)=
\int_{\Omega_1} \mu   J^n_{1,a} \delta \mathbf{F}_1 ( \mathbf{F}^n_{1,a} )^{-1} ( \mathbf{F}^n_{1,a} )^{- \rm T} : \nabla \bv_1 \; {\rm d} \bX_1 - \\
&\int_{\Omega_1} p^{n+1}_{1,k} \text{Cof}(\delta \mathbf{F}_1): \nabla \bv_1 \; {\rm d} \bX_1+
\int_{\Omega_1} J^{n+1}_{1,k} (\mathbf{F}^{n+1}_{1,k} ) ^{- \rm T} : \delta \mathbf{F}_1 q_1 \; {\rm d} \bX_1
- \int_{\Omega_1}  J^{n+1}_{1,k} \delta p_1 ( \mathbf{F}^{n+1}_{1,k} )^{- \rm T} : \nabla \bv_1 \; {\rm d} \bX_1 +  \\
&  \int_{\partial \Omega_C} \delta \bu^{ \rm T}_1 \nabla \mathbf{T}^{n+1}_k \bv_1 \; {\rm d} \Gamma
=-R^{n+1}_{1,k}, \quad \forall \bv_1 \in \bV_{0,h}, q_1 \in Q_{0,h}
\end{aligned}
\end{equation}
where the notation $\delta \bF_1$ stands for $ \nabla (\delta \bu_1) $, and $\text{Cof} (\bF_1)$
is the cofactor operator applied to the deformation gradient $\bF_1$, see Appendix \ref{sec:Appendix}.
For details on the spaces $\bV_{0,h}$ and $Q_{0,h}$, see also Appendix \ref{sec:Appendix}.

Note that the last integral in Eq. \eqref{eq:myo1} is defined on the internal boundary $\partial \Omega_C$ and
accounts for the traction $\mathbf{T}$ acting
on myocyte $\Omega_1$ due to the interaction with myocyte $\Omega_2$.
The Jacobian of the traction vector $\nabla \mathbf{T}$, needed in the linearization, is given by:

\begin{equation}
 \nabla \mathbf{T}=\left( \begin{array}{cc}
\dfrac{\partial T^n}{\partial g_n} & \dfrac{\partial T^n}{\partial g_t}  \\
\dfrac{\partial T^t}{\partial g_n} & \dfrac{\partial T^t}{\partial g_t}  \end{array} \right).
\end{equation}
As a result, the residual $R^{n+1}_{1,k}$ is defined as:

\begin{equation}
\begin{aligned}
  R^{n+1}_{1,k}(\mathbf{u}^{n+1}_{1,k}, p^{n+1}_{1,k})= &  \int_{\Omega_1} \mu   J^n_{1,a} \mathbf{F}^{n+1}_{1,k} ( \mathbf{F}^n_{1,a} )^{-1} ( \mathbf{F}^n_{1,a} )^{- \rm T} : \nabla \bv_1\; {\rm d} \bX_1 -
  \int_{\Omega_1}  J^{n+1}_{1,k} p^{n+1}_{1,k} ( \mathbf{F}^{n+1}_{1,k} )^{- \rm T}  : \nabla \bv_1 \; {\rm d} \bX_1 + \\
& \int_{\Omega_1} (J^{n+1}_{1,k}-1)q_1  \; {\rm d} \bX_1
+ \int_{\partial \Omega_C} \mathbf{T}^{n+1}_k \bv_1 \; {\rm d} \Gamma.
\end{aligned}
\end{equation}
The last integral accounting for the contact between the two myocytes acts
like a traction defined on the Neumann boundary $\partial \Omega_C$.
The linearized system \eqref{eq:myo1} is solved until the incremental norm:
\begin{equation}{\label{eq:incrmec}} {\rm incr_m}=
\dfrac{\Vert \delta \bu_1 \Vert^2_{\mathbf{H}^1(\Omega_1)} }{ \Vert  \bu^{n+1}_{1,k} \Vert^2_{\mathbf{H}^1(\Omega_1)} } +
\dfrac{ \Vert \delta p_1 \Vert^2_{ L^2(\Omega_1)} }{ \Vert  p^{n+1}_ {1,k} \Vert^2_{L^2(\Omega_1)} },
\end{equation}
is less than a prescribed tolerance ${\rm tol_{m}}$.

After convergence, the values of the displacement field $\bu^{n+1}_1$ and pressure $p^{n+1}_1$ inside
myocyte 1 are determined. The updated deformed domain $\Omega^{n+1}_1$ is now given by the coordinates $\bx^{n+1}_1=\bx^{n}_1+ \bu^{n+1}_1$.

\item (E1) Solution of the electrophysiological problem for the myocyte $1$.
Given the values of the electric potential $V^n_1$ and Calcium concentration $w^n_1$, and the same quantities at the
$k$-th iteration of the Newton-Raphson scheme $V^{n+1}_{1,k}$ and $w^{n+1}_{1,k}$ for the myocyte 1,
and the value of the electric potential $V^n_2$ for myocyte 2 at the previous converged iteration, the current
gap in electric potential $\Delta V=\vert V^{n+1,k}_1-V^n_2 \vert$ and the current value $D^{n+1}_{n,k}$ of the
conductivity are computed. The electrophysiological problem
consists in finding the corrections $\delta V_1$ and $\delta w_1$
inside the myocyte $1$, such that for all test functions $\xi_1$ and $\phi_1$ are solutions of
the following linearized reaction-diffusion system:

\begin{subequations}\label{eq:myov1}
 \begin{align}
&\mathcal{E}_{1,V}(V^{n+1}_{1,k}, w^{n+1}_{1,k}; V^{n}_{2}) ( \delta V_1, \delta w_1)=
\dfrac{1}{\Delta t}  \int_{\Omega_1}  \delta V_1 \xi_1 \; {\rm d} \bX_1 +
 \int_{\Omega_1}  (\mathbf{F}^{n+1}_1 )^{-1} \mathbf{D} ( \mathbf{F}^{n+1}_1 )^{- \rm T} \nabla \delta V_1 \cdot \nabla \xi_1 \; {\rm d} \bX_1- \nonumber \\
& \int_{\Omega_1}   \dfrac{\partial I^{n+1}_{k} }{\partial V} \delta V_1 \xi_1 \; {\rm d} \bX - \int_{\Omega_1}   \dfrac{\partial I^{n+1}_{k} }{\partial w} \delta w_1 \xi_1 \; {\rm d} \bX_1 +
\int_{\partial \Omega_C} c \left( \dfrac{\partial D^{n+1}_{n,k}}{\partial V}(V^{n+1}_{1,k}- V^n_2) +D^{n+1}_{n,k}  \right) \delta V_1 \xi_1 \; {\rm d} \Gamma \nonumber \\
&=-R^{n+1,k}_{1,V}  , \quad \forall \xi_1 \in \mathcal{V}_{0,h} \label{eq:myo1v}   \\
&\mathcal{E}_{1,w}(V^{n+1}_{1,k}, w^{n+1}_{1,k}) ( \delta V_1, \delta w_1)=  \dfrac{1}{\Delta t}  \int_{\Omega_1}  \delta w_1 \phi_1 \; {\rm d} \bX_1-
\int_{\Omega_1}\dfrac{ \partial H^{n+1}_{k}}{\partial V} \delta v_1 \phi_1 \; {\rm d} \bX_1-
\int_{\Omega_1} \dfrac{ \partial H^{n+1}_{k}}{\partial w} \delta w_1 \phi_1 \; {\rm d} \bX_1 \nonumber \\
&=  -R^{n+1,k}_{1,w} , \quad \forall  \phi_1 \in \mathcal{W}_{0,h}, \label{eq:myov1w}
 \end{align}
 \end{subequations}
obtained by using the Euler backward time stepping scheme.
The integral defined on $\partial \Omega_C$ in Eq.\eqref{eq:myo1v} represents the linearization of the boundary condition
for electric current across the boundary.
The residual $R^{n+1,k}_{1,V}$ for the equation for $V$ is given by:
\begin{equation*}
\begin{aligned}
R^{n+1,k}_{1,V}=
&\dfrac{1}{\Delta t}  \int_{\Omega_1}  ( V^{n+1}_{1,k}-v^n_1) \xi_1 \; {\rm d} \bX_1+
\int_{\Omega_1}  (\mathbf{F}^{n+1}_1 )^{-1} \mathbf{D}( \mathbf{F}^{n+1}_1 )^{- \rm T} \nabla  V^{n+1}_{1,k}  \cdot \nabla \xi_1 \; {\rm d} \bX_1-
 \int_{\Omega_1}    I^{n+1}_{k}  \xi_1 \; {\rm d} \bX - \\
& \int_{\Omega_1} I_{ \rm app} \xi_1  \; {\rm d} \bX_1 +
\int_{\partial \Omega_C}
c D^{n+1}_{n,k}  (V^{n+1}_{1,k}- V^n_2)    \xi_1 \; {\rm d} \Gamma.
 \end{aligned}
\end{equation*}
Notice that the last integral accounting for the electric contact between the two myocytes, plays the role of an electric flux defined
on the Neumann boundary $\partial \Omega_C$.
The residual $R^{n+1,k}_{1,w}$ for the equation for $w$ is given by:
\begin{equation}
 \begin{aligned}
 R^{n+1,k}_{1,w}=
\dfrac{1}{\Delta t}  \int_{\Omega_1}  ( w^{n+1}_{1,k}-w^n_1) \phi_1\; {\rm d} \bX_1- \int_{\Omega_1}   H^{n+1}_k \phi_1 \; {\rm d} \bX_1 .
 \end{aligned}
\end{equation}
For the derivation of the linearized reaction-diffesion syestem and the finite
element approximating spaces $\mathcal{V}_{0,h}$ and $\mathcal{W}_{0,h}$, see Appendix \ref{sec:Appendix}.

The linearized system \eqref{eq:myov1} is solved untill the norm of the electrophysiological variables
\begin{equation}
\textrm{incr}_e=\Vert \delta V_1 \Vert^2_{L^2(\Omega_1)}+\Vert \delta w_1 \Vert^2_{L^2(\Omega_1)},
 \end{equation}
is less than a prescribed tolerance $ { \rm tol_{\rm e} }$.

\item (M2) Solution of the mechanical problem for the myocyte $2$.
Given the values of
the displacement field vector and pressure $(\bu^{n}_2, p^n_2)$, and the same quantities at the $k$-th iteration of the
Newton-Raphson scheme $(\bu^{n+1}_{2,k}, p^{n+1}_{2,k})$ for the myocyte 2, and
the value of the displacement $\bu^{n+1}_1$ for myocyte 1 calculated in (M1), the current
normal and tangential gaps, $g^{n+1}_{n,k}=(\bu^{n+1}_{2,k}- \bu^{n+1}_1) \cdot \mathbf{n}$
and $g^{n+1}_{t,k}=(\bu^{n+1}_{2,k}- \bu^{n+1}_1) \cdot \mathbf{t}$,
and the cohesive traction vector $\mathbf{T}^{n+1}_k$ are computed. At
each iteration $k+1$, the mechanical problem of contact between the two myocytes
$1$ and $2$ consists in
finding the correction in the displacement vector and
pressure $(\delta \bu_2, \delta p_2)$ inside
myocyte $2$ such that for all
test functions $(\bv_2, q_2)$ are solutions of the linearized system of equations analogous to \eqref{eq:myo1}:

\begin{equation}\label{eq:myo2}
\begin{aligned}
&\mathcal{M}_2(\bu^{n+1}_{2,k}, p^{n+1}_{2,k}; \bu^{n+1}_{1})(\delta \bu_2, \delta p_2)=\int_{\Omega_2} \mu   J^n_{2,a} \delta \mathbf{F}_2 ( \mathbf{F}^n_{2,a} )^{-1} ( \mathbf{F}^n_{2,a} )^{- \rm T} : \nabla \bv_2 \; {\rm d} \bX_2 - \\
&\int_{\Omega_2} p^{n+1}_{2,k} \text{Cof}(\delta \mathbf{F}_2): \nabla \bv_2 \; {\rm d} \bX_2+
\int_{\Omega_2} J^{n+1}_{2,k} (\mathbf{F}^{n+1}_{2,k} ) ^{- \rm T} : \delta \mathbf{F}_2 q_2 \; {\rm d} \bX_2-
 \int_{\Omega_2}  J^{n+1}_{2,k} \delta p_1 ( \mathbf{F}^{n+1}_{2,k} )^{- \rm T} : \nabla \bv_2 \; {\rm d} \bX_2 + \\
&\int_{\partial \Omega_C} \delta \bu^{ \rm T}_2 \nabla \mathbf{T}^{n+1}_k \bv_2 \; {\rm d} \Gamma =-R^{n+1}_{2,k}, \quad \forall \bv_2 \in \bV_{0,h}, q_2 \in Q_{0,h}
\end{aligned}
\end{equation}
where the residual is now given by:
\begin{equation}
\begin{aligned}
  R^{n+1}_{2,k}(\mathbf{u}^{n+1}_{2,k}, p^{n+1}_{2,k})=&
   \int_{\Omega_2} \mu   J^n_{2,a} \mathbf{F}^{n+1}_{2,k} ( \mathbf{F}^n_{2,a} )^{-1} ( \mathbf{F}^n_{2,a} )^{- \rm T} : \nabla \bv_2\; {\rm d} \bX_2 -
   \int_{\Omega_2}  J^{n+1}_{2,k} p^{n+1}_{2,k} ( \mathbf{F}^{n+1}_{2,k} )^{- \rm T}  : \nabla \bv_2 \; {\rm d} \bX_2 + \\
& \int_{\Omega_2} (J^{n+1}_{2,k}-1)q_2  \; {\rm d} \bX_2
+ \int_{\partial \Omega_C} \mathbf{T}^{n+1}_k \bv_2 \; {\rm d} \Gamma.
\end{aligned}
\end{equation}

The resulting system of equations defining the problem (M2) is solved iteratively
until the incremental norm ${\rm \text{incr}_m}$ is less than a given tolerance ${\rm tol_m}$.
The only difference in the
equivalent system of equations for (M2) is the way in which the
gaps and the traction are computed using the solution of the previous problem (M1).
After convergence, the updated (deformed) domain $\Omega^{n+1}_2$ is now given by the coordinates
$\bx^{n+1}_2=\bx^n_2+\bu^{n+1}_2$

\item (E2)  Solution of the electrophysiological problem for the myocyte $2$.
Given the values of the electric potential $V^n_2$ and Calcium concentration $w^n_2$, and the same quantities at the
$k$-th iteration of the Newton-Raphson scheme $V^{n+1}_{2,k}$ and $w^{n+1}_{2,k}$ for the myocyte 2,
and the value of the electric potential $V^{n+1}_1$ for myocyte 1 computed as solution of the problem (E1), the current
gap in electric potential $\Delta V=\vert V^{n+1,k}_2-V^{n+1}_1 \vert$ and the current value $D^{n+1}_{n,k}$ of the
conductivity are computed. The electrophysiological problem (E2)
consists in finding the corrections $\delta V_2$ and $\delta w_2$
inside the myocyte $2$, such that for all test functions $\xi_2$ and $\phi_2$ are solutions of
the following linearized reaction-diffusion system:

\begin{subequations}\label{eq:myov2}
 \begin{align}
&\mathcal{E}_{2,V}(V^{n+1}_{2,k}, w^{n+1}_{2,k}; V^{n+1}_{1})( \delta V_2, \delta w_2)=
\dfrac{1}{\Delta t}  \int_{\Omega_2}  \delta V_2 \xi_2 \; {\rm d} \bX_2 +
 \int_{\Omega_2}  (\mathbf{F}^{n+1}_2 )^{-1} \mathbf{D} ( \mathbf{F}^{n+1}_2 )^{- \rm T} \nabla \delta V_2 \cdot \nabla \xi_2 \; {\rm d} \bX_2- \nonumber \\
& \int_{\Omega_2}   \dfrac{\partial I^{n+1}_{k} }{\partial V} \delta V_2 \xi_2 \; {\rm d} \bX_2 - \int_{\Omega_2}   \dfrac{\partial I^{n+1}_{k} }{\partial w} \delta w_2 \xi_2 \; {\rm d} \bX_2 +
\int_{\partial \Omega_C} c \left( \dfrac{\partial D^{n+1}_{n,k}}{\partial V}(V^{n+1}_{2,k}- V^{n+1}_1) +D^{n+1}_{n,k}  \right) \delta V_2 \xi_2 \; {\rm d} \Gamma \nonumber \\
&=-R^{n+1,k}_{2,V}  , \quad \forall \xi_2 \in \mathcal{V}_{0,h} \label{eq:myo2v}   \\
&\mathcal{E}_{2,w}(V^{n+1}_{2,k}, w^{n+1}_{2,k})( \delta V_2, \delta w_2)=
\dfrac{1}{\Delta t}  \int_{\Omega_2}  \delta w_2 \phi_2 \; {\rm d} \bX_2-
\int_{\Omega_2}\dfrac{ \partial H^{n+1}_{k}}{\partial V} \delta V_2 \phi_2 \; {\rm d} \bX_2-  \int_{\Omega_2} \dfrac{ \partial H^{n+1}_{k}}{\partial w} \delta w_2 \phi_2 \; {\rm d} \bX_2 \nonumber \\
&=  -R^{n+1,k}_{2,w} , \quad \forall  \phi_2 \in \mathcal{W}_{0,h}, \label{eq:myov1w}
 \end{align}
 \end{subequations}
where the residual $R^{n+1,k}_{2,V}$ for the equation for $V$ is given by:
\begin{equation*}
\begin{aligned}
 R^{n+1,k}_{2,V}=
&\dfrac{1}{\Delta t}  \int_{\Omega_2}  ( V^{n+1}_{2,k}-v^n_2) \xi_2 \; {\rm d} \bX_2+
 \int_{\Omega_2}  (\mathbf{F}^{n+1}_2 )^{-1} \mathbf{D}( \mathbf{F}^{n+1}_2 )^{- \rm T} \nabla  V^{n+1}_{2,k}  \cdot \nabla \xi_2 \; {\rm d} \bX_2-
 \int_{\Omega_2}    I^{n+1}_{k}  \xi_2 \; {\rm d} \bX_2
+ \\
&\int_{\partial \Omega_C} c D^{n+1}_{n,k}  (V^{n+1}_{2,k}- V^{n+1}_1)    \xi_2 \; {\rm d} \Gamma.
 \end{aligned}
\end{equation*}
and the residual $R^{n+1,k}_{2,w}$ for the equation for $w_2$ is given by:
\begin{equation}
 \begin{aligned}
 R^{n+1,k}_{2,w}=\dfrac{1}{\Delta t}  \int_{\Omega_2}  ( w^{n+1}_{2,k}-w^n_2) \phi_2\; {\rm d} \bX_2- \int_{\Omega_2}   H^{n+1}_k \phi_2 \; {\rm d} \bX_2 .
 \end{aligned}
\end{equation}

The resulting system of equations defining problem (E2) is solved iteratively
until the incremental norm ${\rm \text{incr}_e}$ is less than a given tolerance ${\rm tol_e}$.
The only difference in the equivalent system of equations for (E2) is the way in which the transjunctional gaps
and the conductivities are computed using the solution of the previous problem (E1).
\end{enumerate}
At each timestep, the described sequence of subproblems (M1), (E1), (M2) and (E2) is iterated until
convergence of the physical interface conditions defined on $\partial \Omega_C$:
\begin{subequations}\label{eq:physical}
\begin{align}
	&\dfrac{| g_n^{n+1} - g_n^{n+1,l} |}{| g_n^{n+1} |} < {\rm tol}_p \,,\quad
	g_n^{n+1,l+1} \leftarrow g_n^{n+1} \,,
	\\
	&\dfrac{| g_t^{n+1} - g_t^{n+1,l} |}{| g_t^{n+1} |} < {\rm tol}_p \,,\quad
	g_t^{n+1,l+1} \leftarrow g_t^{n+1} \,,
	\\
	&\dfrac{| \Delta V^{n+1} - \Delta V^{n+1,l} |}{| \Delta V^{n+1} |} < {\rm tol}_p \,,\quad
	\Delta V^{n+1,l+1} \leftarrow \Delta V^{n+1} \,,
\end{align}
\end{subequations}
where tol$_p=10^{-5}$.
This physical interface condition is denoted as (P).
The algorithm for the proposed time integration with a staggered
scheme is detailed in Algorithm \ref{alg1}.

\begin{algorithm*}[ht!]
 \textbf{Input mechanical parameters:} $\mu, \mathbf{a}_l, \mathbf{a}_t, T_{n, \rm max}, g_{n, \rm max}, f, a_t, \alpha$;  \\
 \textbf{Input electrophysiological parameters:} $a, b, c_1, c_2, d, D_l, D_t, I_{\rm app}, a_1, a_2, a_3, a_4, c $ ; \\
  \textbf{Input activation:} $\beta_c, c_0, c^*, g^{\rm max}_0, \epsilon$ ;\\
  \For{$n=1, \dots , N_{ \rm it}$ $\mathrm{time \ steps}$}{
   \While{the physical conditions at the interface (P) are not satisfied }{
      Compute $\bu^{n+1}_1$ and $p^{n+1}_1$ solving the mechanical subproblem (M1) with value at the iterface $\bu^n_2$ \;
  \While{$(   \mathrm{incr_m} \ge \mathrm{tol}_m)$}{
    $ \mathcal{M}_1(\bu^{n+1}_{1,k}, p^{n+1}_{1,k}; \bu^{n}_{2})(\delta \bu_1, \delta p_1)=-R^{n+1}_{1,k}$ \;
     $\bu^{n+1}_{1,k+1} \leftarrow \bu^{n+1}_{1,k} + \delta \bu^{n+1}_1 $ ,  $p^{n+1}_{1,k+1} \leftarrow p^{n+1}_{1,k} + \delta p^{n+1}_1 $\;
   }{
   Compute $V^{n+1}_1$ and $w^{n+1}_1$ solving the electrophysiological subproblem (E1) with value at the iterface $V^n_2$ \;
       \While{$(   \mathrm{incr}_e    \ge \mathrm{tol}_e)$}{
   \begin{equation*}
   \begin{cases}
     \mathcal{E}_{1,V}(V^{n+1}_{1,k}, w^{n+1}_{1,k}; V^{n}_{2})(\delta V_1, \delta w_1)=-R^{n+1}_{1,V} \\
     \mathcal{E}_{1,w}(V^{n+1}_{1,k}, w^{n+1}_{1,k})(\delta V_1, \delta w_1)=-R^{n+1}_{1,w}
     \end{cases}
      \end{equation*}
      $V^{n+1}_{1,k+1} \leftarrow V^{n+1}_{1,k} + \delta V^{n+1}_1 $ ,  $w^{n+1}_{1,k+1} \leftarrow w^{n+1}_{1,k} + \delta w^{n+1}_1 $\;
   }
   Compute $\bu^{n+1}_2$ and $p^{n+1}_2$ solving the mechanical subproblem (M2) with value at the iterface $\bu^{n+1}_1$ \;
     \While{$(   \mathrm{incr_m} \ge \mathrm{tol}_m)$}{
$ \mathcal{M}_2(\bu^{n+1}_{2,k}, p^{n+1}_{2,k}; \bu^{n+1}_{1})(\delta \bu_2, \delta p_2)=-R^{n+1}_{2,k}$ \;
     $\bu^{n+1}_{2,k+1} \leftarrow \bu^{n+1}_{2,k} + \delta \bu^{n+1}_2 $ ,  $p^{n+1}_{2,k+1} \leftarrow p^{n+1}_{2,k} + \delta p^{n+1}_2 $\;
   }
   Compute $V^{n+1}_2$ and $w^{n+1}_2$ solving the electrophysiological subproblem (E2) with value at the iterface $V^{n+1}_1$ \;
   \While{$(   \mathrm{incr}_e    \ge \mathrm{tol}_e)$}{
      \begin{equation*}
   \begin{cases}
     \mathcal{E}_{2,V}(V^{n+1}_{2,k}, w^{n+1}_{2,k}; V^{n+1}_{1})(\delta V_2, \delta w_2)=-R^{n+1}_{2,V} \\
      \mathcal{E}_{2,w}(V^{n+1}_{2,k}, w^{n+1}_{2,k})(\delta V_2, \delta w_2)=-R^{n+1}_{2,w}
      \end{cases}
     \end{equation*}
      $V^{n+1}_{2,k+1} \leftarrow V^{n+1}_{2,k} + \delta V^{n+1}_2 $ ,  $w^{n+1}_{2,k+1} \leftarrow w^{n+1}_{2,k} + \delta w^{n+1}_2 $\;
   }
   }
     }
 }
 \caption{Numerical scheme for the solution of the structure-structure interaction problem for the electromechanical contact of two cardiomyocytes.}
 \label{alg1}
\end{algorithm*}

\section{Numerical experiments}
\label{sec:NumericalApplications}

Several numerical examples are herein conducted in order to investigate the mechanical and electro-physiological responses of the interaction problem between cardiomyocytes for different test geometries and
different values of the model parameters. Action potential diffusivities are scaled to the myocytes length scale such to reproduce a slow Calcium wave propagating behavior as in~\cite{ruiz:2014}, i.e. $D_l=D_t=0.06$ $\mu$m$^2$/ms.

\subsection{Myocyte geometry model}
\label{ssec:Geometry}
\begin{figure*}[ht!]
\begin{center}
    \subfigure[AR1: $52 \,\mu m \times 26 \,\mu m$; $\partial\Omega_C = 24\,\mu m$]
    {\includegraphics[height=0.135\textwidth]{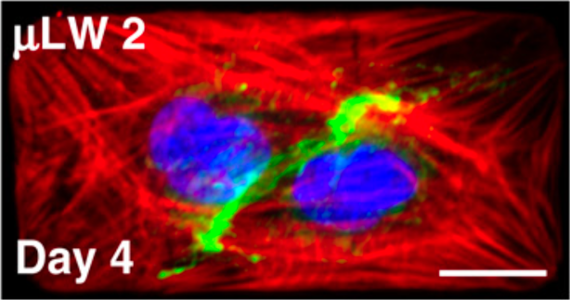}
    \includegraphics[trim=0cm 2.3cm 0cm 0cm, clip=true, totalheight=0.135\textwidth, angle=0]{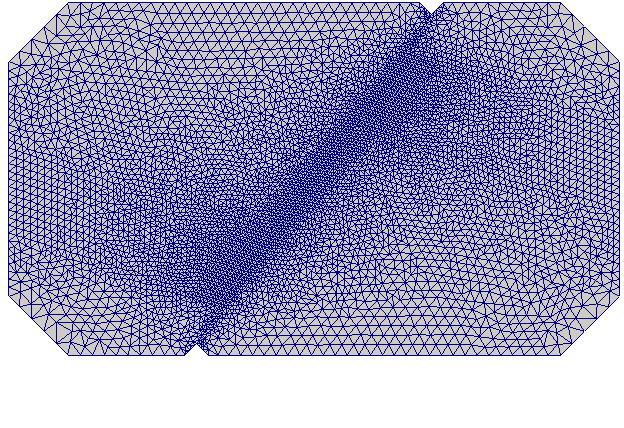}
    }
    \\
     \subfigure[AR2: $72 \,\mu m \times 18 \,\mu m$; $\partial\Omega_C = 26\,\mu m$]
    {\includegraphics[height=0.105\textwidth]{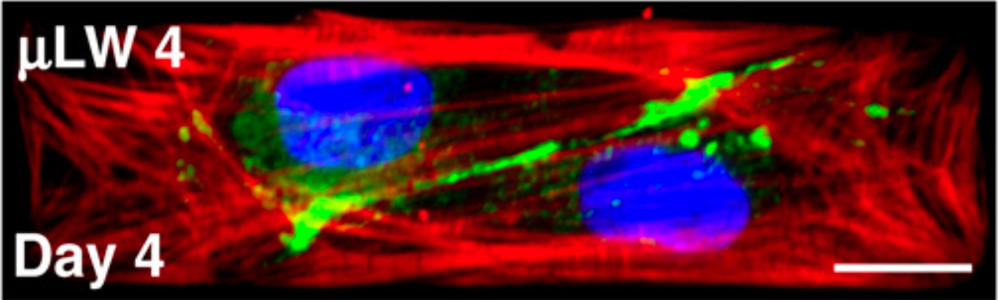}
     \includegraphics[trim=0cm 2.3cm 0cm 0cm, clip=true, totalheight=0.105\textwidth, angle=0]{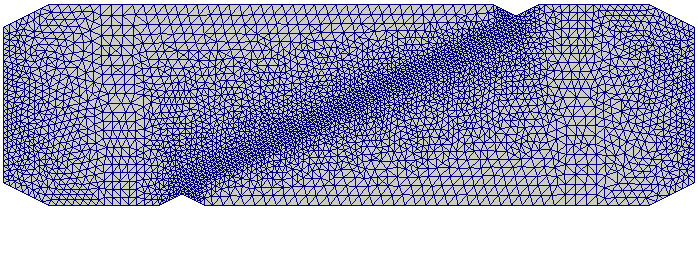}
    }
    \\
     \subfigure[AR3: $112 \,\mu m \times 14 \,\mu m$; $\partial\Omega_C = 38\,\mu m$]
    {\includegraphics[height=0.084\textwidth]{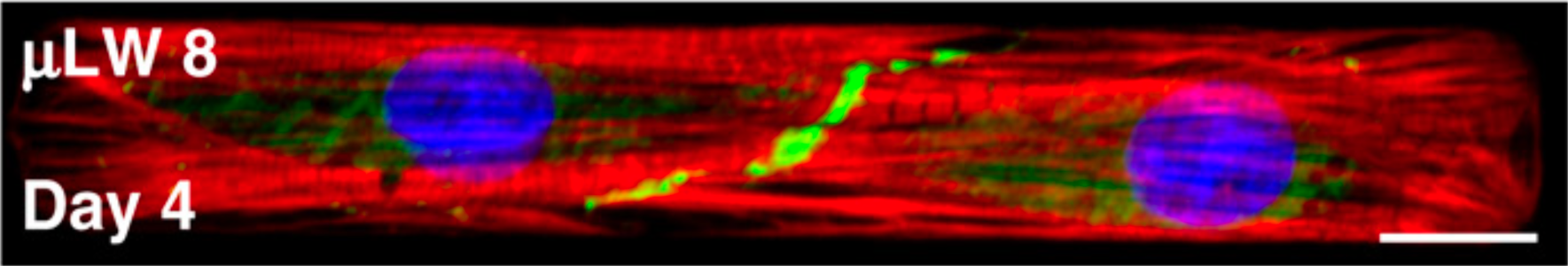}
    \includegraphics[trim=0cm 2.3cm 0cm 0cm, clip=true, totalheight=0.084\textwidth, angle=0]{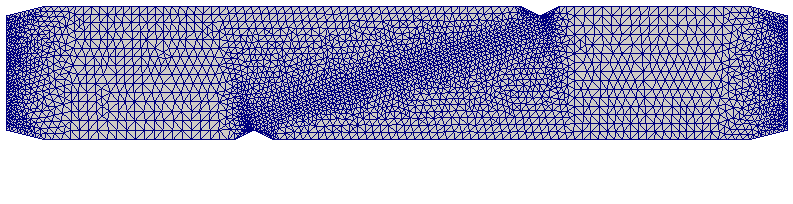}
    }
    \caption{
    (Left) In vitro $\mu$-engineered cardiomyocyte cell-cell configurations at four days of differentiation (reproduction with permission from McCain et al. PNAS 109:9881--9886, 2012 \cite{mccain:2012}).
    (Right) Test geometries corresponding to the three aspect ratios (a) AR1, (b) AR2 and (c) AR3, respectively.}
    \label{fig:Geom}
\end{center}
\end{figure*}

The geometry, boundary conditions and mesh discretization adopted for the numerical simulations are
provided in Fig.~\ref{fig:Geom}. The vectors defining the fibers orientations are chosen as the unit vectors of the orthonormal basis of $\mathbb{R}^2$, i.e.,
$\ba_l=\be_1$ and $\ba_t=\be_2$.
With reference to Fig.~\ref{fig:mesh}, let $L$ and $H$ be, respectively, the sizes of the domain $\Omega_1 \cup \Omega_2$.
Myocytes are assumed to have the same material parameters.

An electrical stimulus $I_{\rm app}=\exp(-10(X^2_1+(X_2-H/2)^2))$ is applied for a time interval $[0,2]$ ms to  myocyte 1 on a patch near its left boundary, centered in its middle position.
This stimulation elicits the onset of an action potential wave propagating from myocyte 1 to myocyte 2 passing through the interface $\partial \Omega_C$.
In order to asses how the inclination of the interface $\partial \Omega_C$ with respect to the orientation of the longitudinal fiber vectors $\ba_l$ affects the propagation of electrical waves, we consider different values of
$L$ and $H$. Correspondingly, we define three aspect ratios AR1, AR2, AR3 defined by the ratio $L/H=52/26, 72/18, 112/14$ $[\mu{\rm m}/\mu{\rm m}]$, as in  experimental $\mu$-engineered cell cultures~\cite{mccain:2012}.
Accordingly, three lengths of the interface $\partial \Omega_C$ are obtained:
$24, 26$, and $38 \; \mu \rm m$ for AR1, AR2 and AR3, respectively.
These three geometries corresponding to their own aspect ratios shown in Fig.~ \ref{fig:Geom} are observed in different temporal stages of cardiomyocyte differentiation, namely at four days of tissue development.
In all the numerical simulations a refinement of the mesh is adopted on the interface boundary $\partial \Omega_C$.
Unstructured triangular meshes are considered containing
$11308, 9320, 6034$ elements for AR1, AR2 and AR3, respectively.
The total simulation time was set $T_{\rm fin}=1100 \; \rm ms$ with a time step $\Delta t=1 \; \rm ms$.

\subsection{The role of the myocytes' stiffness}
\begin{figure}[ht!]
\begin{center}
      {\includegraphics
      [trim=9.5cm 0cm 13cm 1cm, clip=true, totalheight=0.3\textwidth, angle=0]{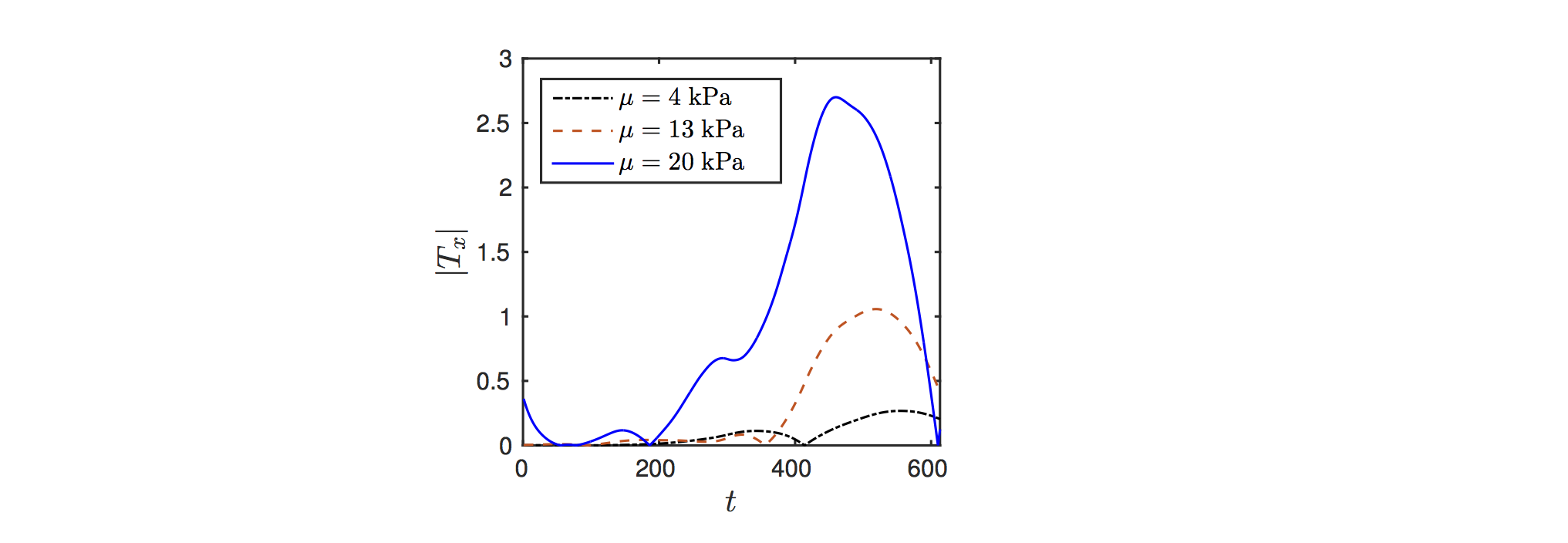}}
    \caption{Absolute value of the horizontal traction component $\vert T_x \vert$ during myocyte contraction for
    low $(\mu=4 \rm kPa)$, moderate $(\mu=13 \rm kPa)$ and high $(\mu= 20\rm kPa)$ elastic modulus.}
    \label{fig:sigmavsT}
\end{center}
\end{figure}

The first test is conducted to assess the role of the myocytes' stiffness by changing the value of $\mu$. Considering a test geometry described by AR1, three situations are examined corresponding to
low $(\mu=4 \rm kPa)$,
moderate $(\mu=13 \rm kPa)$, and
high $(\mu=20 \rm kPa)$ elastic modulus.
Interface properties are $c=0.1$ and $T_{n,{\rm max}}=1000$ kPa.
Since our model is two-dimensional, the elastic modulus of the cell can be interpreted as a parameter to model the stiffness of the substrate to which the cell  adheres.
Accordingly, Fig.~\ref{fig:sigmavsT} shows the evolution of the absolute value of the horizontal traction $\vert T_x \vert$ evaluated in the central point of the interface boundary $\partial \Omega_C$. Simulation results are in good agreement with experimental trends~\cite{mccain:2012} both in shape and amplitude. In particular, by increasing $\mu$, higher stresses are induced (about one order of magnitude higher as shown in Fig. S3 in \cite{mccain:2012}). In addition, the time course of the computed traction $\vert T_x \vert$ resembles the morphology shown in the experimental traces. The model is able also to detect small
oscillations due to the nonlinearity of the action potential wave, i.e. sharp propagating front, plateau and the smooth repolarization phase.

\subsection{The role of the interface mechanical properties}
\begin{figure}[ht!]
\centering
\includegraphics[trim=9cm 0cm 10cm 0cm, clip=true, totalheight=0.3\textwidth, angle=0]{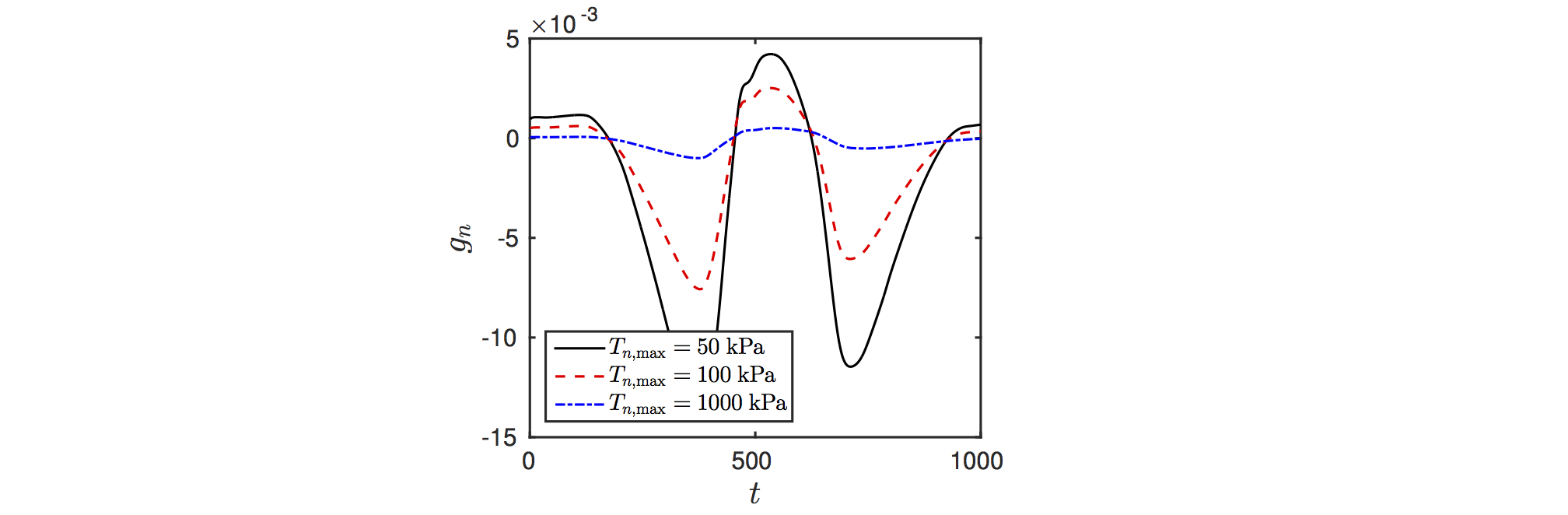}
\caption{Evolution of the normal gap $g_n$ at the central point of the interface for a maximum traction $T_{n, \rm max}= \{ 50, 100, 1000 \}$ kPa.
\label{fig.gap}}
\end{figure}

In this test, considering again the AR1 geometry and setting $\mu=4$ kPa, the maximum adhesive traction is varied as $T_{n, \rm max}=\{ 50, 100, 1000 \} \; \rm kPa$, keeping $g_{n,{\rm max}}=1$ and $c=0.1$. Accordingly, we compute the normal gap $g_n$ at the central point of the
boundary $\partial \Omega_C$ as shown in Fig.~\ref{fig.gap}. As expected, when the value of the maximum traction increases, the normal gap decreases, and, in addition, non-negligible oscillations of $g_n$ are computed due to the crossing of the excitation wave on the interface.
The value of $T_{n, \rm max}=1000 \; \rm kPa$ is in perfect agreement with the usual material properties of cells~\cite{deshpande:2006} and is considered appropriate to reproduce the experimental observations~\cite{mccain:2012}. In particular, a strong physiological cell-cell adhesion is obtained. Lower cell-cell- adhesion could be responsible for pathological situations as again highlighted in~\cite{mccain:2012}.

\subsection{The role of the interface electrical conductivity and of the myocyte aspect ratio}
We conclude our analysis with a series of numerical simulations of the electromechanical contact problem considering $\mu=4$ kPa and the interface mechanical parameters identified in the previous subsection.
Different values of the corrective term $c=\{0.01, 0.1, 0.5 \}$ are investigated modifying the conductivity $D_n$ at the interface boundary $\partial \Omega_C$.
Moreover, the effect of the myocytes aspect ratio is also analyzed.

\begin{figure*}[htp!]
\centering
\subfigure[AR1]{
\includegraphics[trim=8.5cm 0cm 12.1cm 0cm, clip=true, totalheight=0.25\textwidth, angle=0]{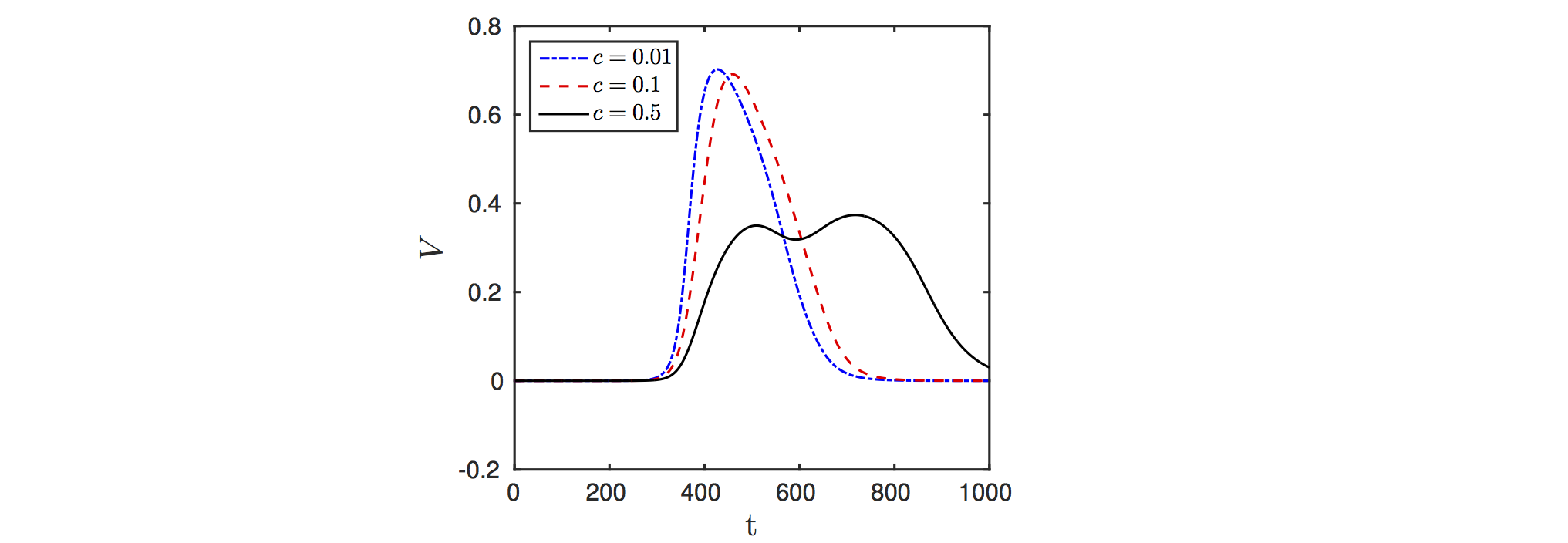}
\includegraphics[trim=8.5cm 0cm 12.1cm 0cm, clip=true, totalheight=0.25\textwidth, angle=0]{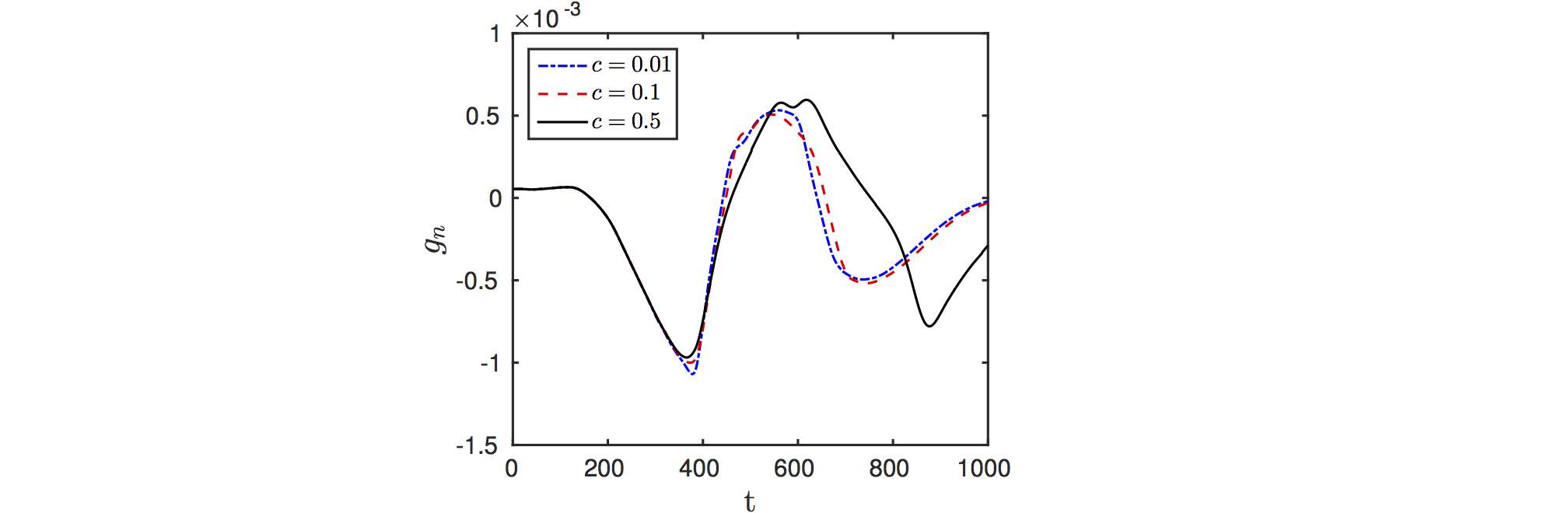}
\includegraphics[trim=8.5cm 0cm 12.1cm 0cm, clip=true, totalheight=0.25\textwidth, angle=0]{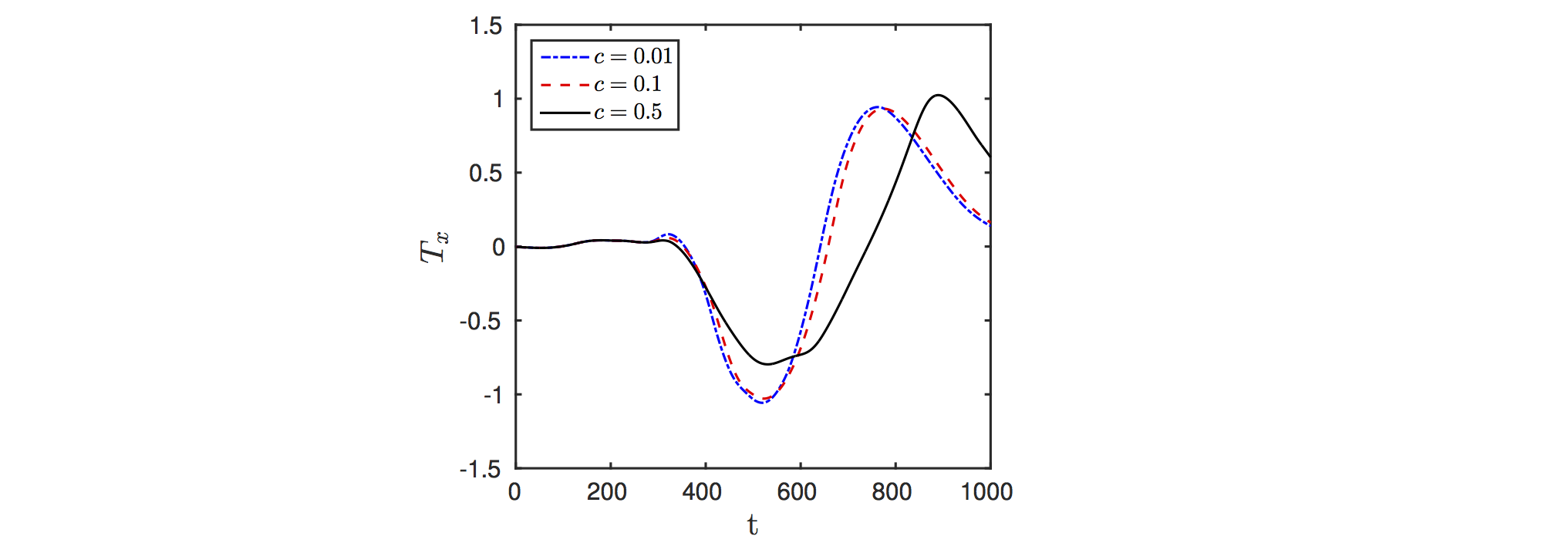}
}
\\
\subfigure[AR2]{
\includegraphics[trim=8.5cm 0cm 12.1cm 0cm, clip=true, totalheight=0.25\textwidth, angle=0]{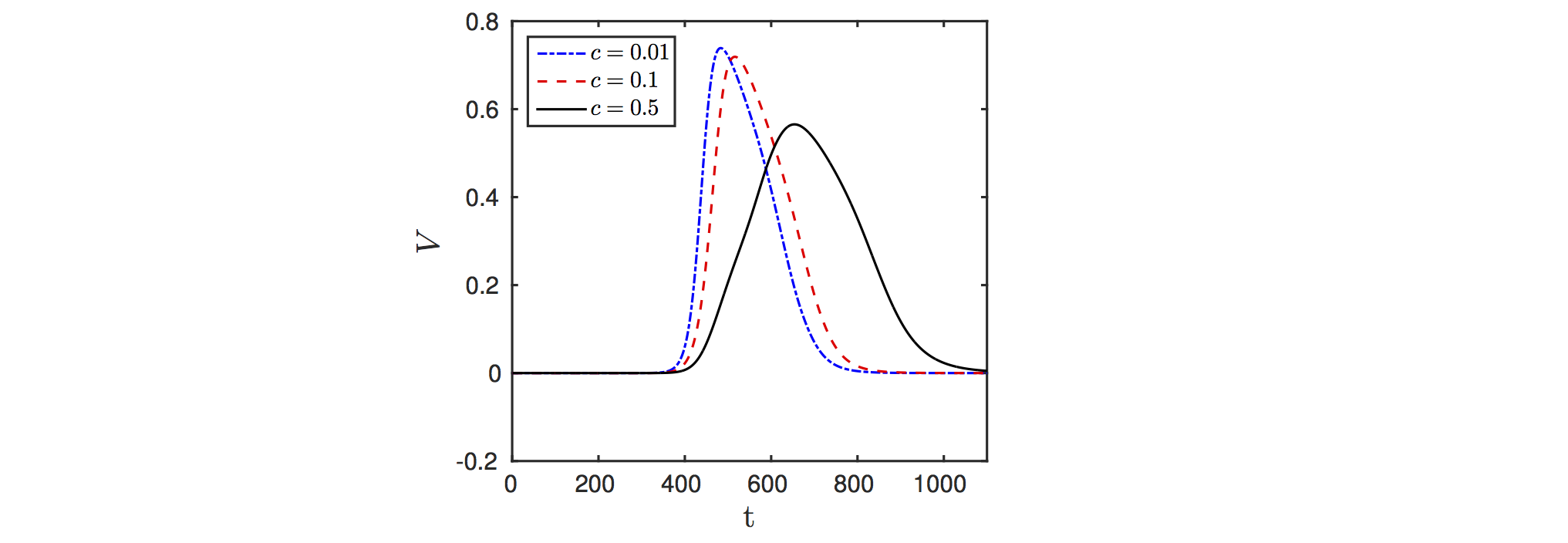}
\includegraphics[trim=8.5cm 0cm 12.1cm 0cm, clip=true, totalheight=0.25\textwidth, angle=0]{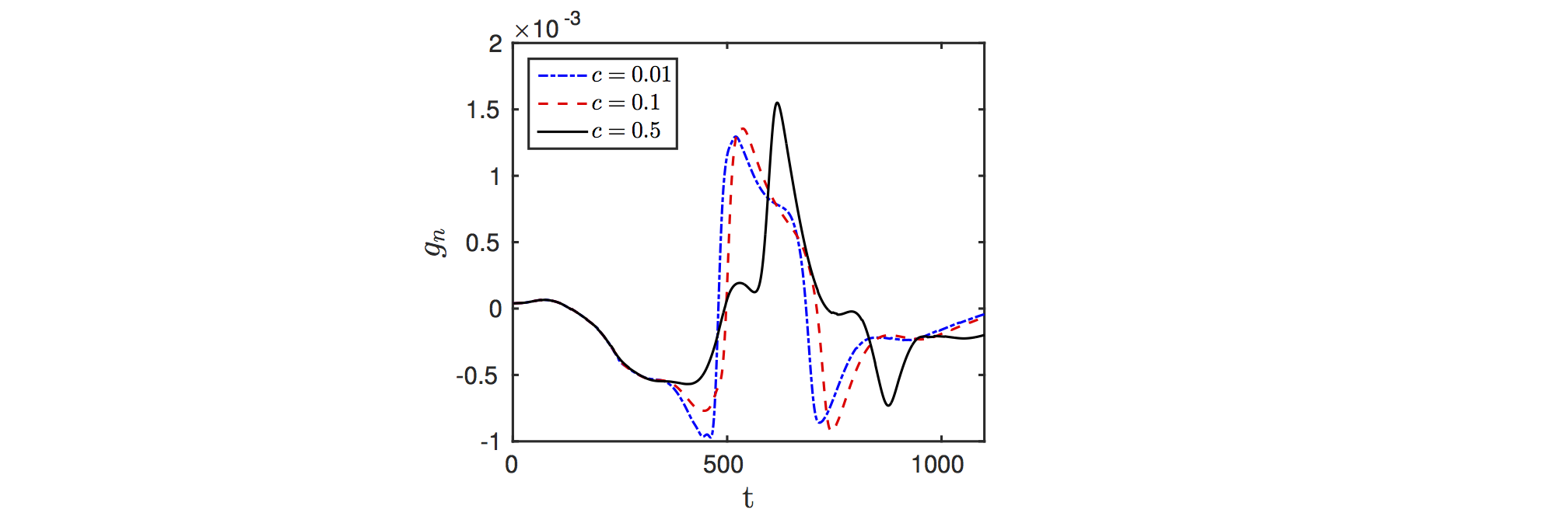}
\includegraphics[trim=8.5cm 0cm 12.1cm 0cm, clip=true, totalheight=0.25\textwidth, angle=0]{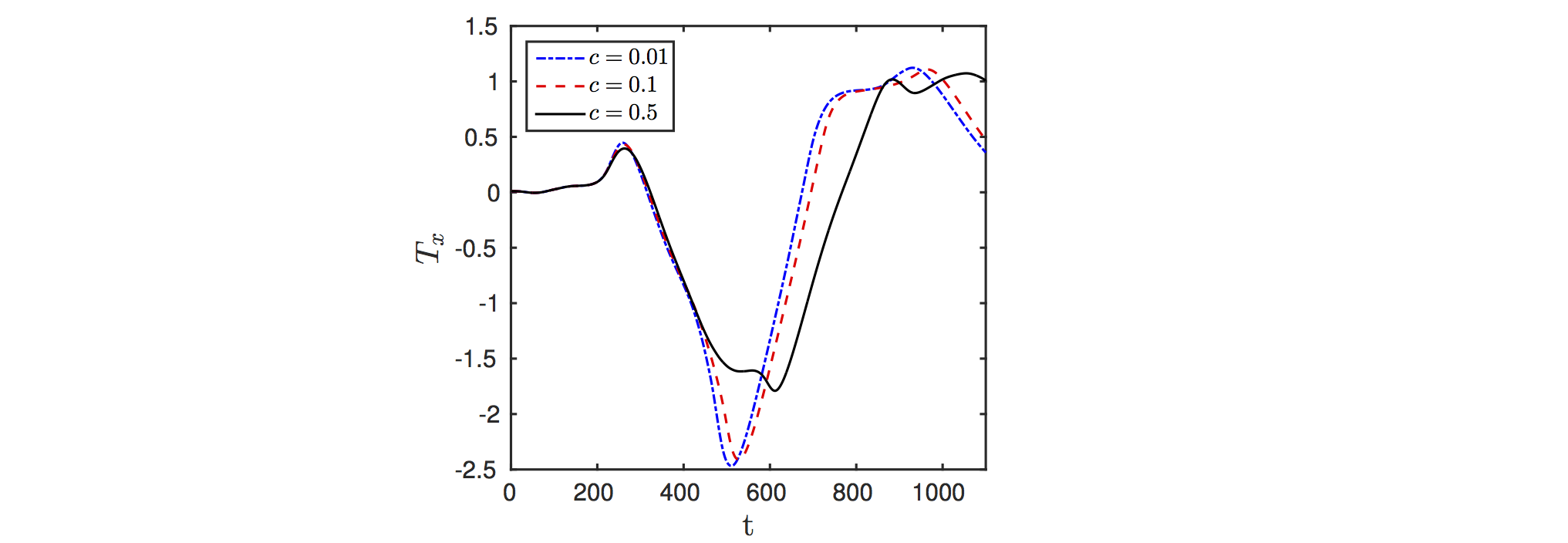}
}
\\
\subfigure[AR3]{
\includegraphics[trim=8.5cm 0cm 12.1cm 0cm, clip=true, totalheight=0.25\textwidth, angle=0]{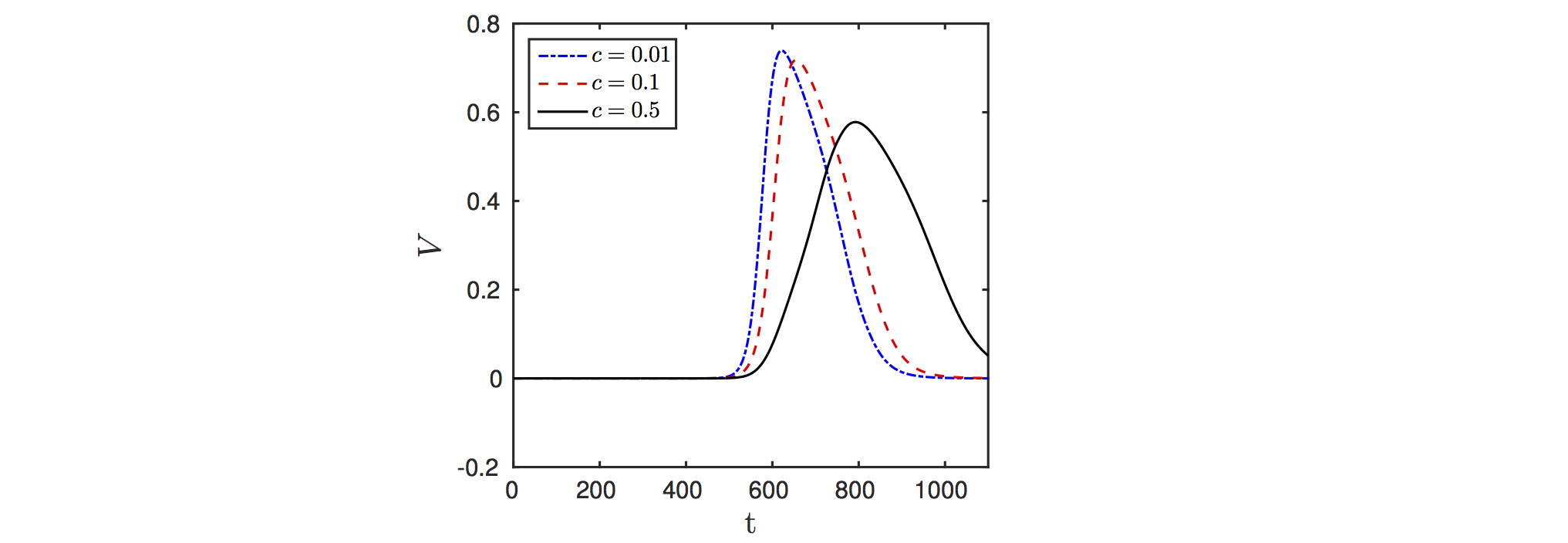}
\includegraphics[trim=8.5cm 0cm 12.1cm 0cm, clip=true, totalheight=0.25\textwidth, angle=0]{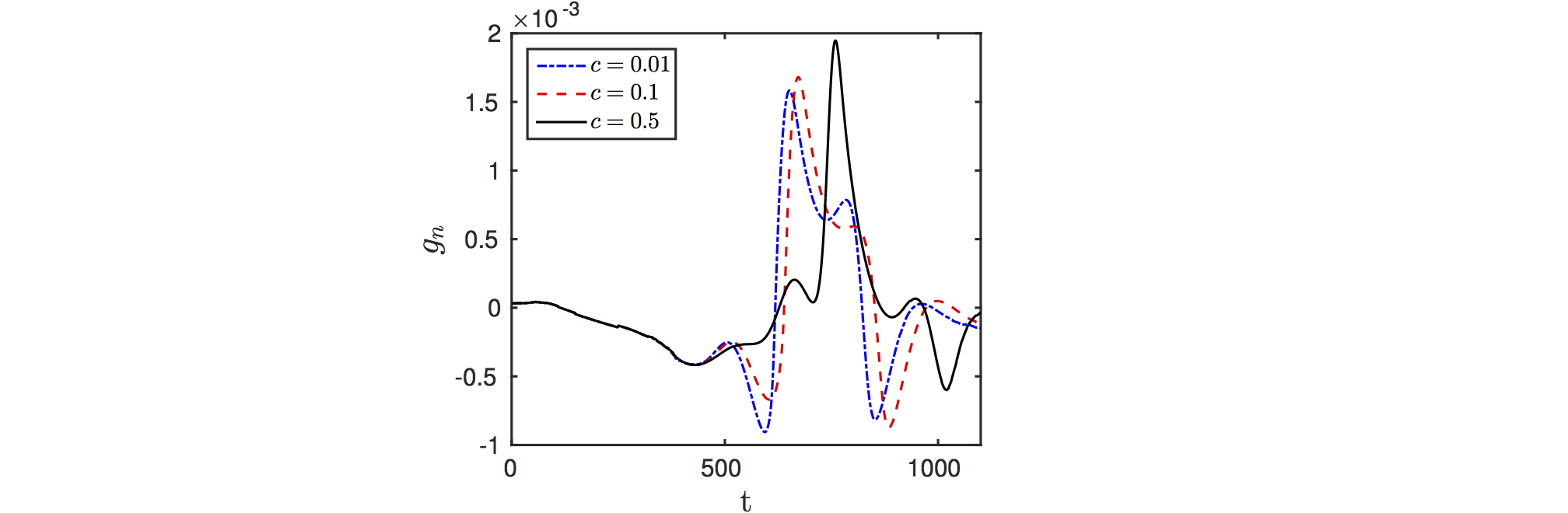}
\includegraphics[trim=8.5cm 0cm 12.1cm 0cm, clip=true, totalheight=0.25\textwidth, angle=0]{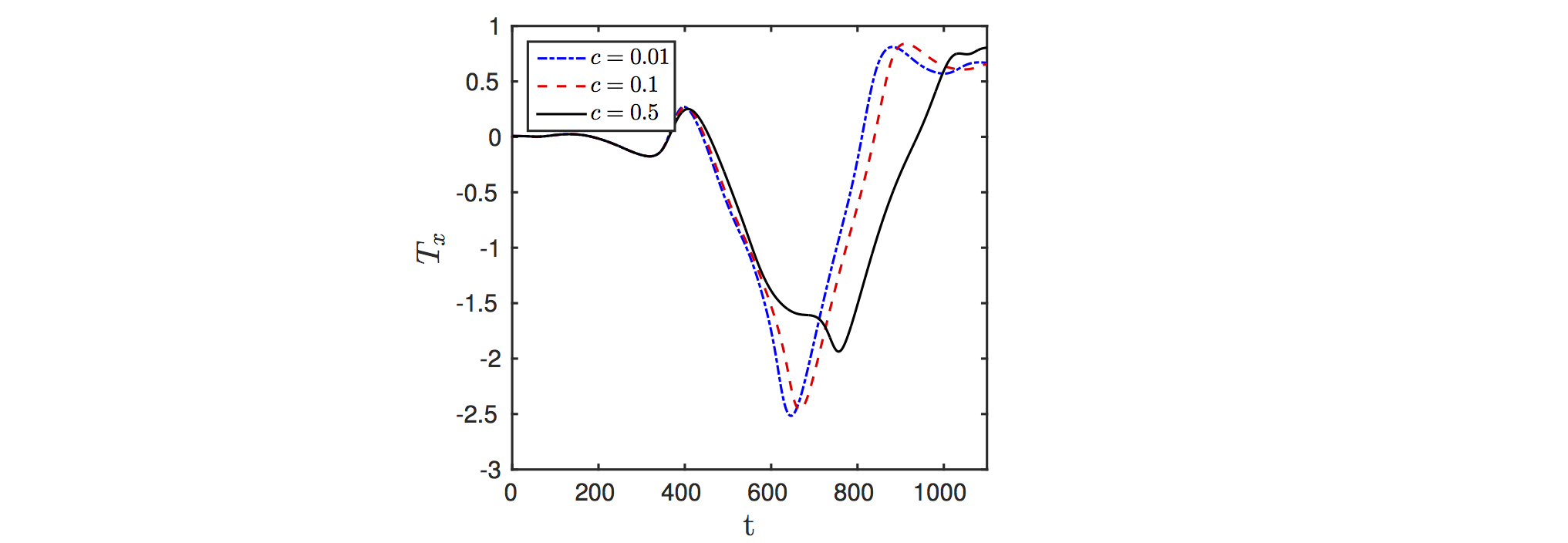}
}
\caption{Time evolution of $V$ in myocyte 1 at the interface, $g_n$, and $T_x$ for (a) AR1, (b) AR2 and (c) AR3 for $c=\{0.01, 0.1, 0.5 \}$.}
\label{fig:AR123A}
\end{figure*}

Figure~\ref{fig:AR123A} shows the time evolution of the interface dimensionless
voltage $V$,
normal gap $g_n$, and
horizontal traction $T_x$ computed
in the central point of the interface $\partial \Omega_C$.
Low values of the corrective coefficient, e.g. $c=0.01$, characterize a \emph{zero flux} electrical boundary. This condition, usually applied in cardiac reaction diffusion models~\cite{cherry:2011}, mimics an open boundary allowing the voltage wave to travel from myocyte 1 to myocyte 2 without obstacles.
As the value of $c$ reaches $0.1$ then a small shift of the observed quantities is gained though strong additional nonlinear effects appear for $c=0.5$. In particular, this last condition induces multiple unexpected oscillations in all the computed fields. The normal gap varies within $10^{-3}$ values in all the cases and the expected scaling between $T_n$ and $g_n$ is recovered (not shown).
The traction component in the x direction $T_x$, on the other hand, resembles the propagation direction. $T_x$, in fact, assumes negative values when the action potential wave $V$ travels within myocyte 1 while it reverses and becomes positive when the wave enters myocyte 2. Such a behavior is due to the experienced sequences of contraction and relaxation induced by the active dynamics over the whole domain $\Omega_1\cup\Omega_2$.

The extended comparison of these distinct behaviors visualizing the structure-structure contact problem is provided in Figs.~\ref{fig:AR1frame},
\ref{fig:AR2frame}, \ref{fig:AR3frame} (see Supplementary Material for representative videos). Selected frames of the deformed configuration are shown during the action potential propagation across the interface boundary. These numerical analyses confirm that the combined effects of nonlinear diffusivity and aspect ratio can induce non-negligible deviations of the excitation wave when crossing the cell-cell interface. Accordingly, eventual wave breaks may arise from this condition which may represent one of the key factors in the onset of cardiac alternans and arrhythmias~\cite{gizzi:2013}.

\begin{figure*}[htp!]
\centering
\subfigure[AR1]{\includegraphics[trim=8.5cm 0cm 12.1cm 0cm, clip=true, totalheight=0.25\textwidth, angle=0]{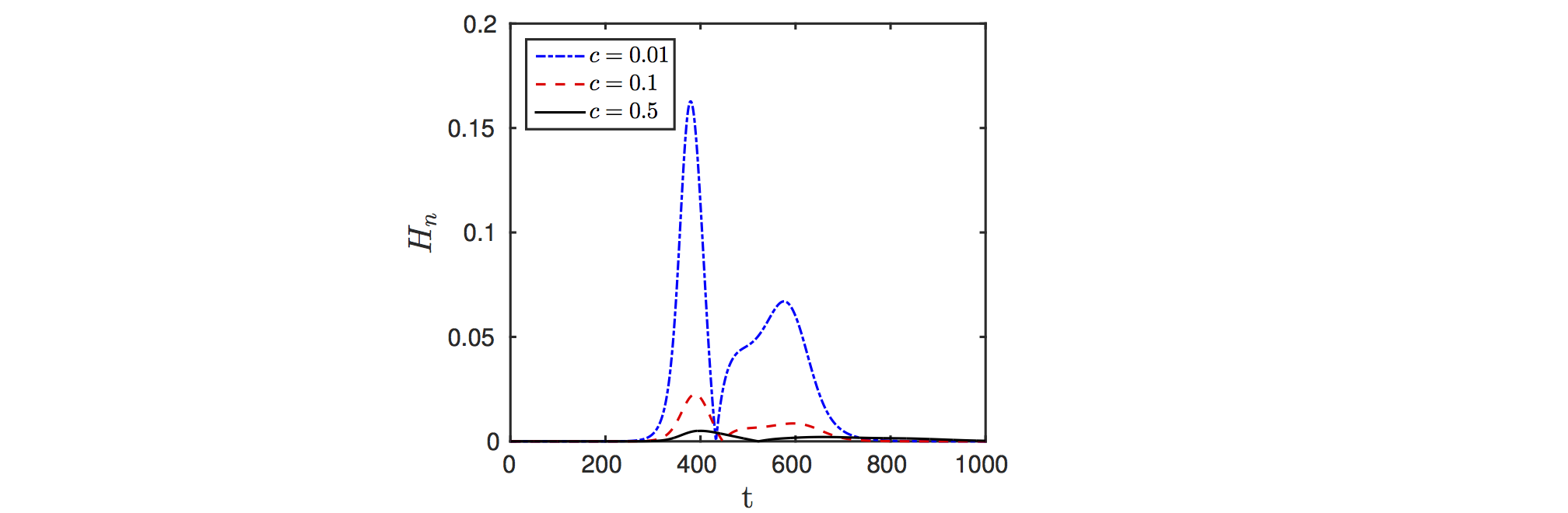}}
\subfigure[AR2]{\includegraphics[trim=8.5cm 0cm 12.1cm 0cm, clip=true, totalheight=0.25\textwidth, angle=0]{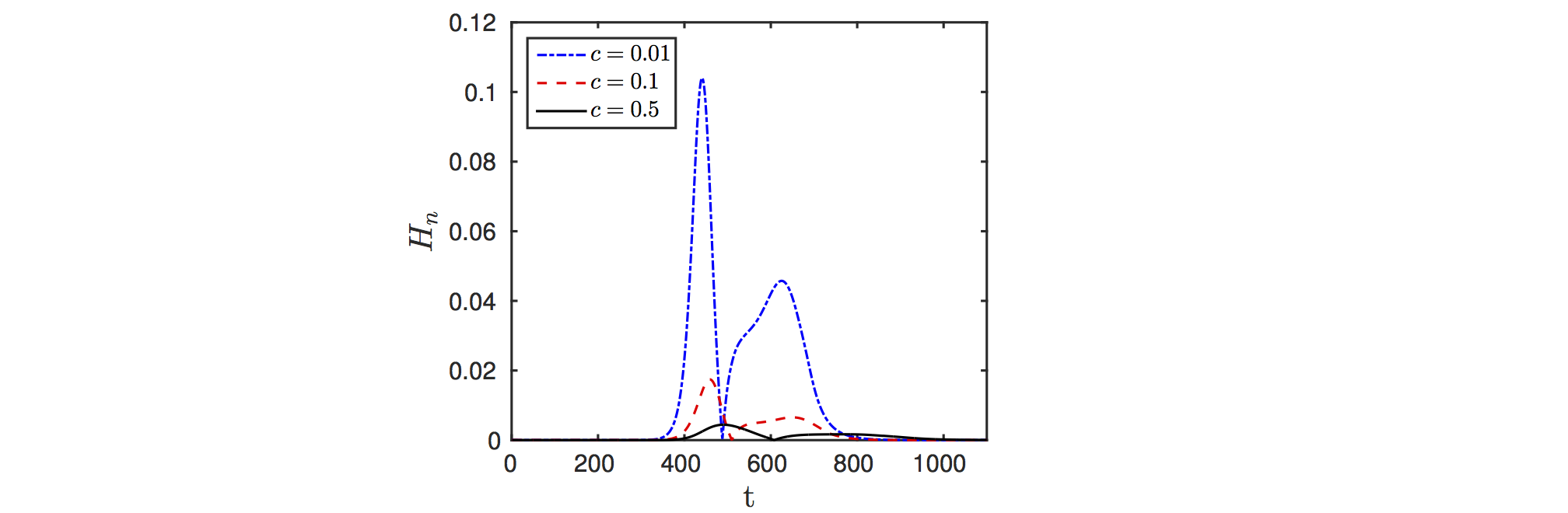}}
\subfigure[AR3]{\includegraphics[trim=8.5cm 0cm 12.1cm 0cm, clip=true, totalheight=0.25\textwidth, angle=0]{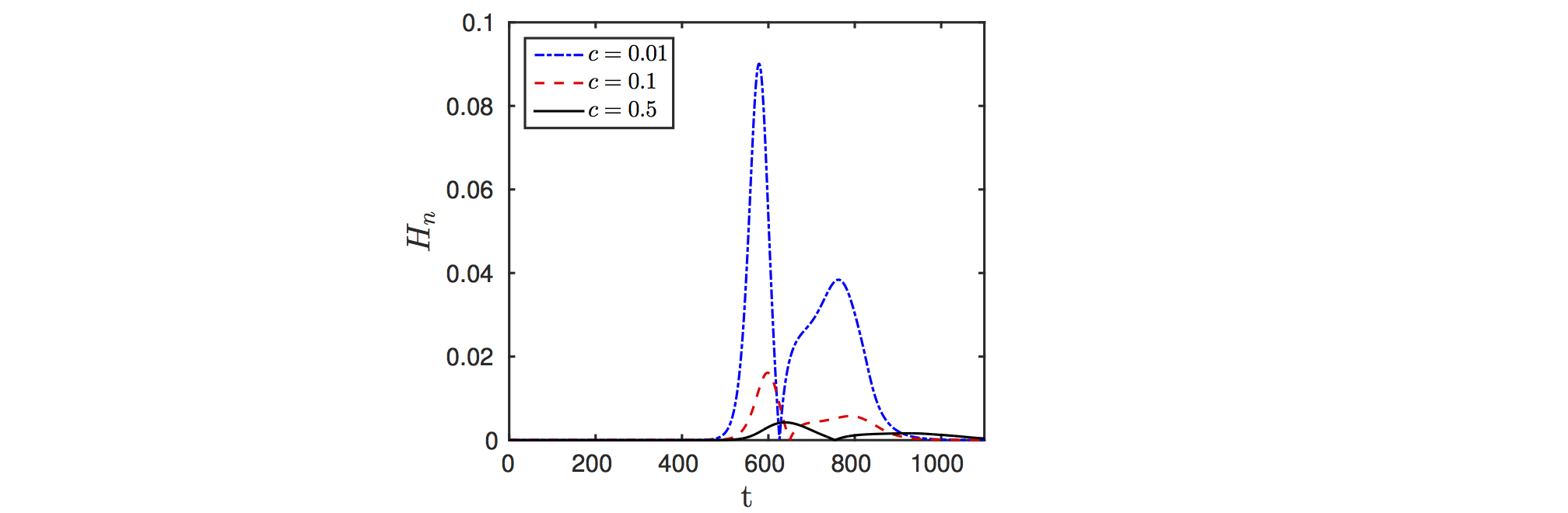}}
\caption{Evolution of the normal interface flux $H_n$ in time for (a) AR1, (b) AR2) and (c) AR3 for $c=\{0.01, 0.1, 0.5 \}$.}
\label{fig:AR123B}
\end{figure*}

As additional level of analysis, we further quantify the nonlinear electrical flux $H_n=\nabla \left( \bD \nabla V \right) \cdot \mathbf{n}$ exchanged on the interface  between the two cells. This particular information, which is one of the novel ingredients resulting from the electromechanical contact problem, is strictly related with the information transmission properties.
As expected from theoretical and physiological arguments, the shape and morphology of the normal interface electrical flux does not change as the aspect ratio is varied.
However, major variations are observed for the different values of the coefficient $c$ inducing a higher/smaller levels of the flux. Moreover, a slight increase in the amplitude of the flux is quantified for the smallest aspect ratio AR1 due to the fact that the amount of energy delivered on the tissue is the same for the three cases but the flux is distributed on a shorter interface. Such an energetic reasoning would be of fundamental importance for the study of emergent dynamics in clusters of cardiomycytes~\cite{mertz:2012}.

\section{Conclusions}
\label{sec:Conclusions}

In this work we proposed a novel structure-structure computational modeling approach for the numerical simulation of nonlinear contact problems arising in soft biological tissues. We formulated an extended electromechanical model accounting for a multi-field interface description of the nonlinear electric fluxes and adhesive contact mechanics properties between cardiomyocytes. Our approach is framed within a staggered finite element solution strategy allowing us to reproduce the main physical features of the cell-cell system, and opening the possibility to scale up the simulations to a large number of myocytes thanks to the inherent parallelization. The proposed multi-field contact problem matches several experimental results both qualitatively and quantitatively and further allows to compute interface fluxes of fundamental importance for the correct information transmission between the two cells.

The present contribution paves the root for a number of theoretical generalizations and engineering applications. The framework can be easily extended to more realistic electrophysiological models as well as incorporate generalized hyperelastic material models. At the same time, high performance computing studies can be based on this work considering clusters of cardiomyocytes interacting according to the features depicted here and reproducing emerging phenomena typical of excitable biological tissues.

\section*{Acknoledgments}
Authors would like to thank the European Research Council for supporting the ERC Starting Grant ``Multi-field and multi-scale Computational Approach to Design and Durability of PhotoVoltaic Modules'' - CA2PVM, under the European Union's Seventh Framework Programme (FP/2007- 2013)/ERC Grant Agreement n. 306622.

\bibliographystyle{unsrt}      

\newpage

\begin{figure*}[htp!]
\centering
\includegraphics[width=.32\textwidth]{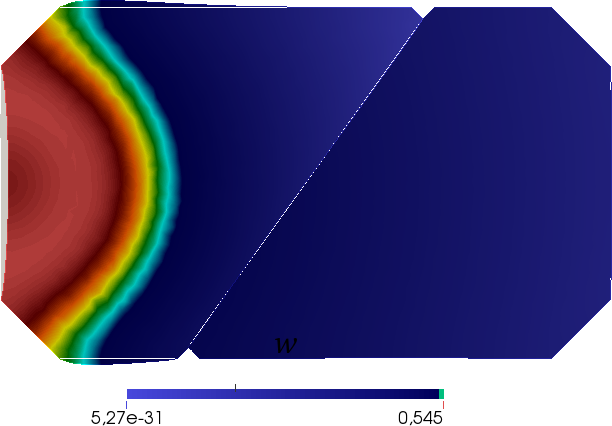}
\includegraphics[width=.32\textwidth]{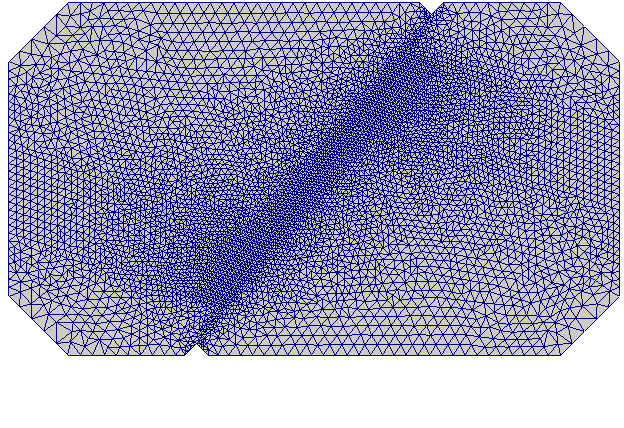} \\
\includegraphics[width=.32\textwidth]{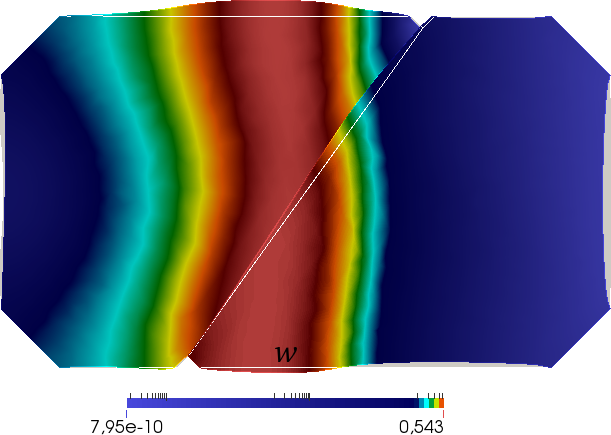}
\includegraphics[width=.32\textwidth]{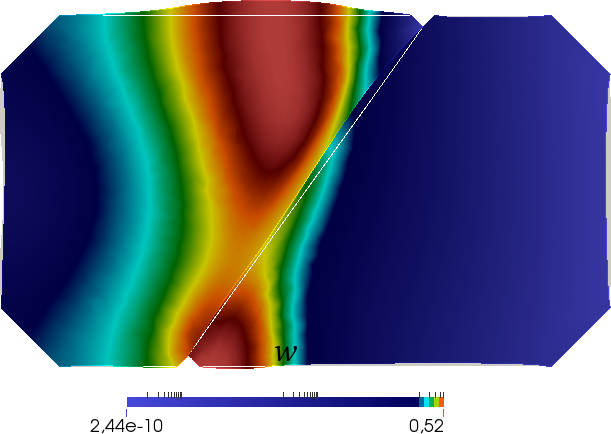} \\
\includegraphics[width=.32\textwidth]{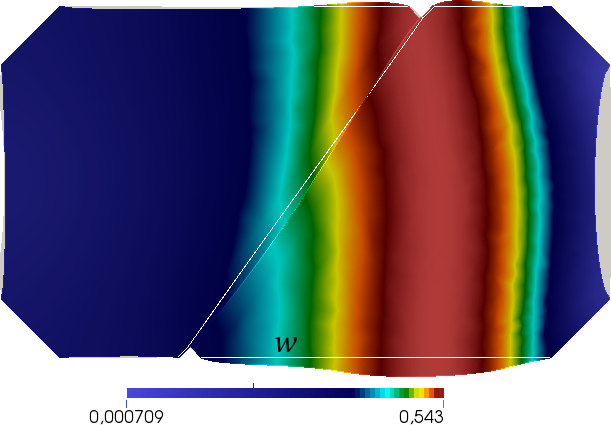}
\includegraphics[width=.32\textwidth]{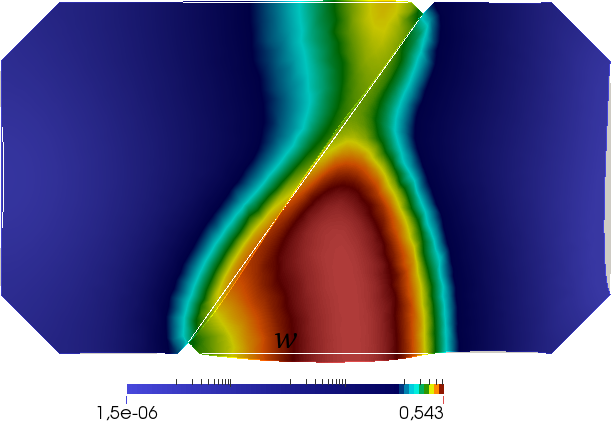} \\
\includegraphics[width=.32\textwidth]{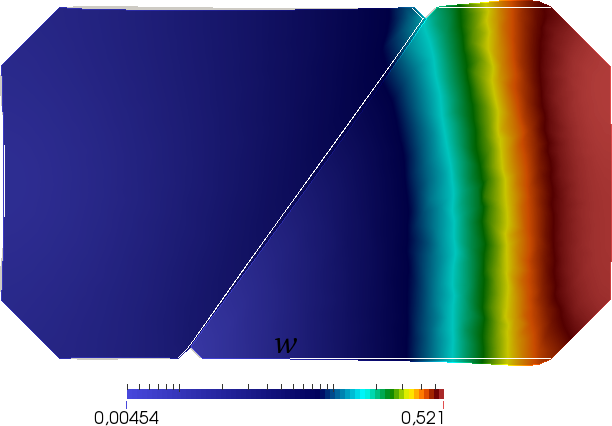}
\includegraphics[width=.32\textwidth]{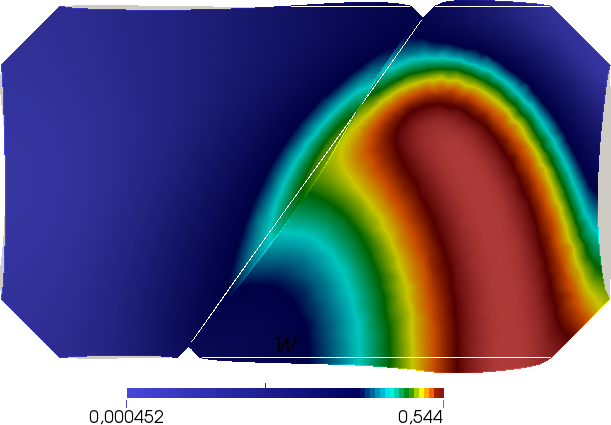} \\
\subfigure[$c=0.01$]{\includegraphics[width=.32\textwidth]{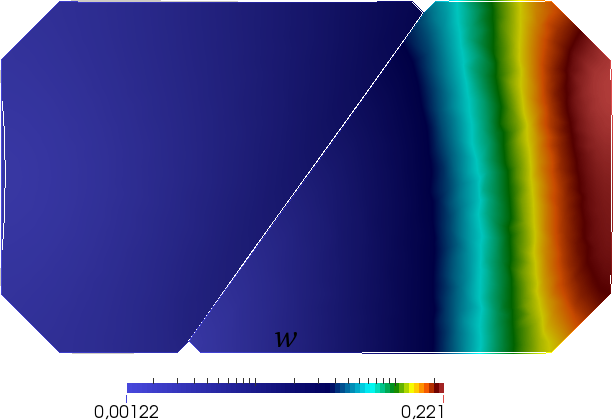} }
\subfigure[$c=0.5$]{\includegraphics[width=.32\textwidth]{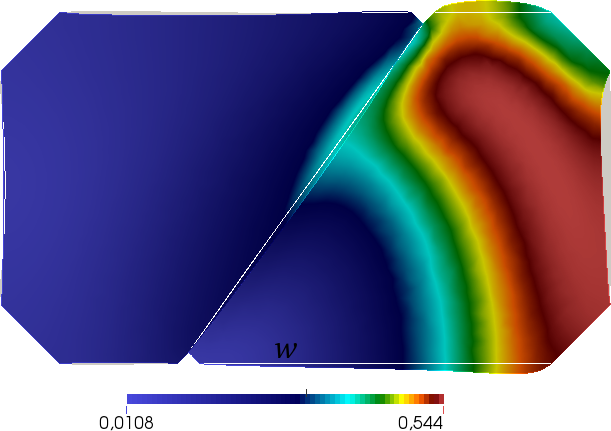}}
\caption{AR1 geometry and mesh. Time evolution of deformed domain and of the Calcium concentration wave $w$ for $c=\{0.01, 0.5 \}$.}
\label{fig:AR1frame}
\end{figure*}

\begin{figure*}[htp!]
\centering
\includegraphics[width=.48\textwidth]{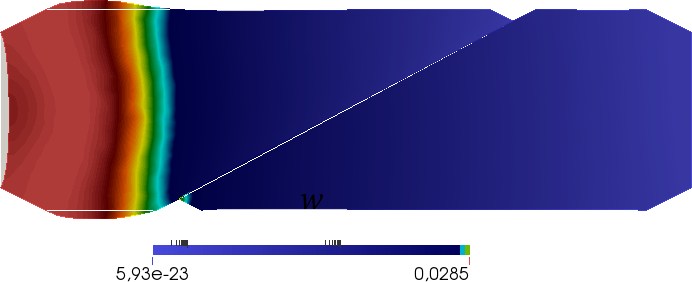}
\includegraphics[width=.48\textwidth]{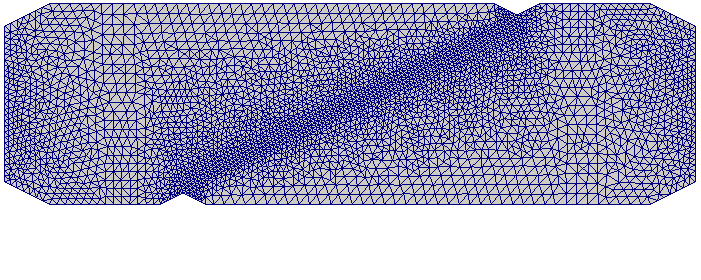} \\
\includegraphics[width=.48\textwidth]{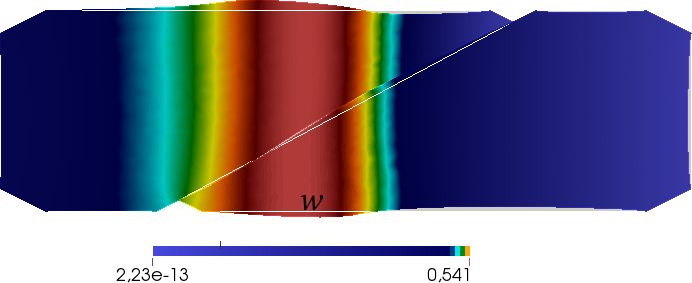}
\includegraphics[width=.48\textwidth]{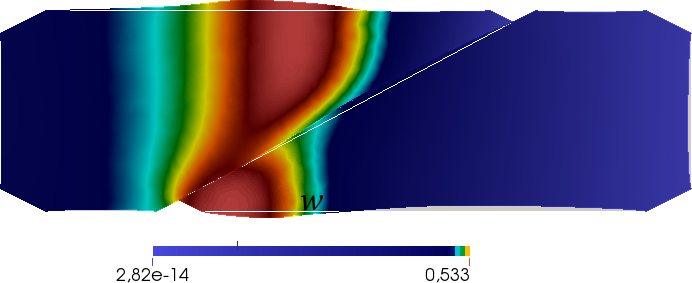} \\
\includegraphics[width=.48\textwidth]{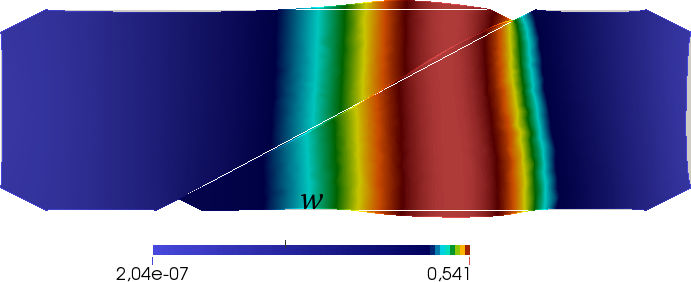}
\includegraphics[width=.48\textwidth]{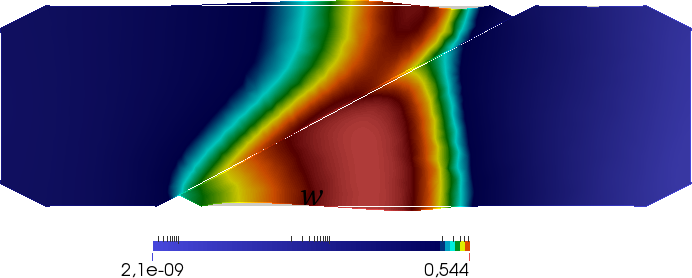} \\
\includegraphics[width=.48\textwidth]{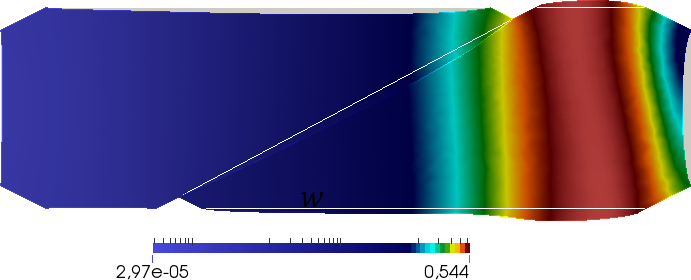}
\includegraphics[width=.48\textwidth]{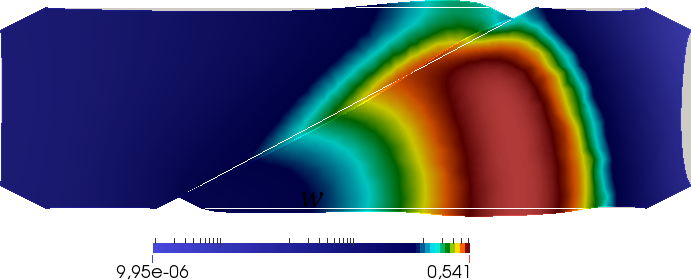} \\
\subfigure[$c=0.01$]{\includegraphics[width=.48\textwidth]{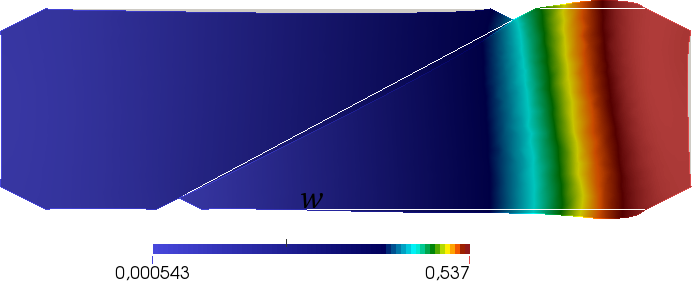}}
\subfigure[$c=0.5$]{\includegraphics[width=.48\textwidth]{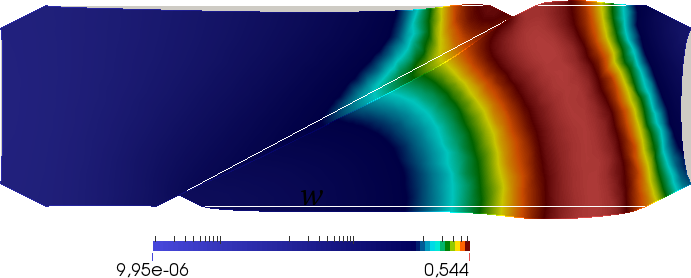}}
\caption{AR2 geometry and mesh. Time evolution of deformed domain and of the Calcium concentration wave $w$ for $c=\{0.01, 0.5 \}$.}
\label{fig:AR2frame}
\end{figure*}

\begin{figure*}[htp!]
\centering
\includegraphics[width=.49\textwidth]{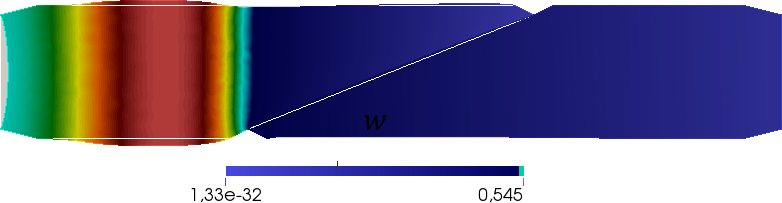}
\includegraphics[width=.49\textwidth]{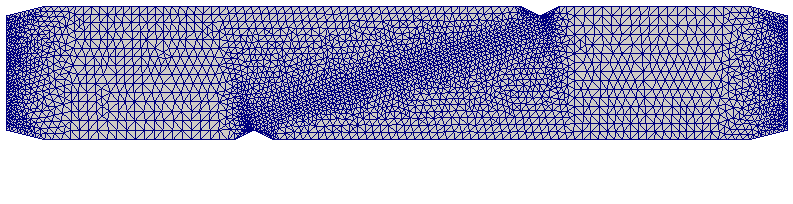} \\
\includegraphics[width=.49\textwidth]{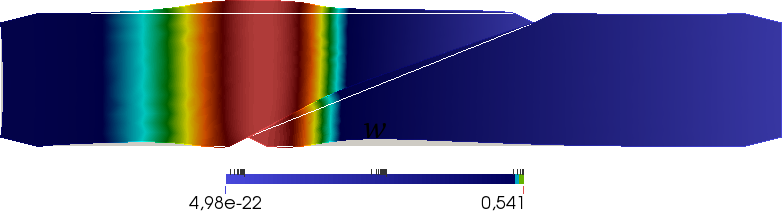}
\includegraphics[width=.49\textwidth]{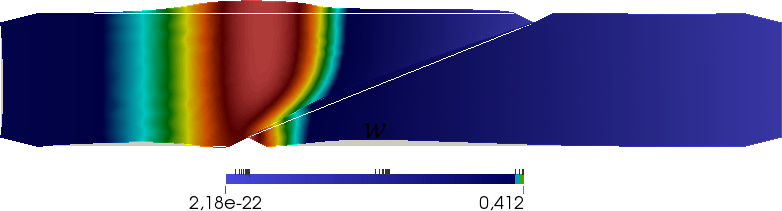} \\
\includegraphics[width=.49\textwidth]{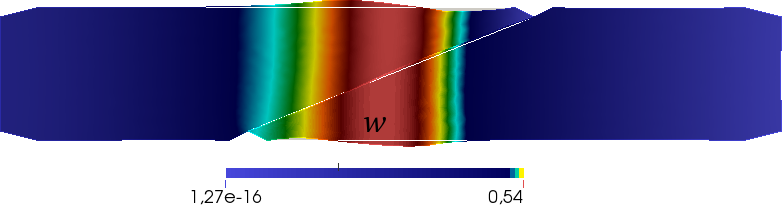}
\includegraphics[width=.49\textwidth]{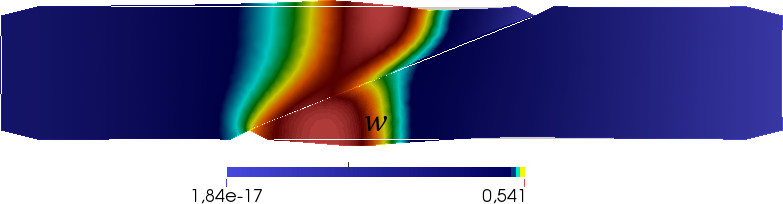} \\
\includegraphics[width=.49\textwidth]{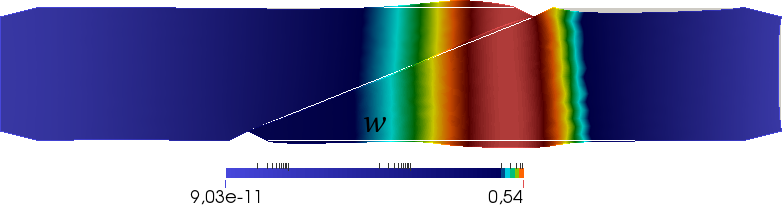}
\includegraphics[width=.49\textwidth]{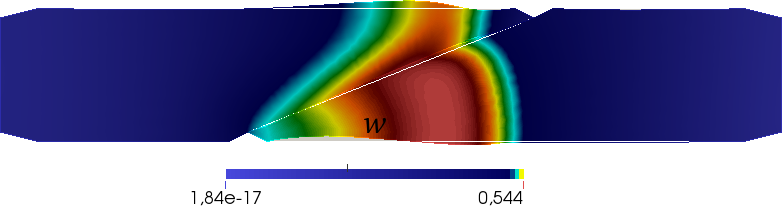} \\
\subfigure[$c=0.01$]{\includegraphics[width=.49\textwidth]{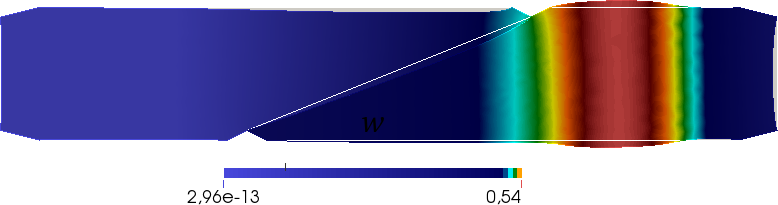}}
\subfigure[$c=0.5$]{\includegraphics[width=.49\textwidth]{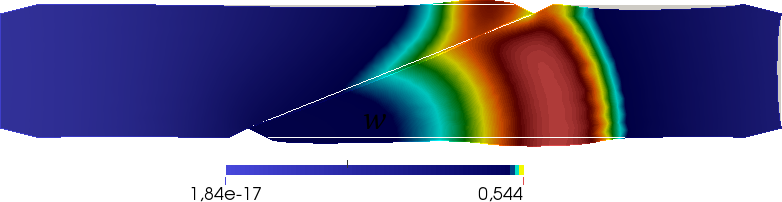}}
\caption{AR3 geometry and mesh. Time evolution of deformed domain and of the Calcium concentration wave $w$ for $c=\{0.01, 0.5 \}$.}
\label{fig:AR3frame}
\end{figure*}

\begin{appendix}

\section{Finite element discretization and staggered solution scheme}
\label{sec:Appendix}
The staggered solution scheme for the solution of the coupled electromechanical problem for a single myocyete is derived.
The displacement vector $\bu$ is approximated in the finite element vector space $\bV_h=[ \mathbb{P}^2(\Omega_0) ]^2$
of continuous piecewise quadratic triangular Lagrange finite elements, the pressure $p$ is approximated in the space $Q_h=\mathbb{P}^1(\Omega_0)$ of linear triangular finite elements, while
the electric potential $V$ and the Calcium concentration $w$ are approximated in the spaces of $\mathcal{V}_h=\mathcal{W}_h=\mathbb{P}^2(\Omega_0)$
of quadratic triangular finite elements. The time interval is partitioned into $t^0=0 \dots \leq t^n \leq  \dots t^N=T_{\rm fin}$ and $t^{n+1}=t^n+ \Delta t$ where $\Delta t$ is the timestep.
At each time step $t^{n+1}$ the mechanical problem \eqref{eq:Wstress} is solved first, where
$\mathbf{F}^n_a$ and $J^n_a$ are evaluated at the previous values $\gamma^n_l$, $\gamma^n_t$, such that the following problem is solved
for all test functions $\bv, q$:
\begin{equation}\label{eq:mech}
 \begin{aligned}
 &  \int_{\Omega_0} \mu  J^n_a \mathbf{F}^{n+1} ( \mathbf{F}^n_a )^{-1} ( \mathbf{F}^n_a )^{- \rm T} : \nabla \bv  \; {\rm d} \bX-
\int_{\Omega_0}   J^{n+1} p^{n+1} ( \mathbf{F}^{n+1} )^{- \rm T} : \nabla \bv \; {\rm d} \bX=0, \quad \forall \bv \in \bV_{0,h}, \\
&\int_{\Omega_0} (J^{n+1}-1)q \; {\rm d} \bX  = 0, \quad \forall q \in Q_{0,h},
 \end{aligned}
\end{equation}
where the superscript ${(\cdot)}^{n+1}$ stands for the evaluation on the current field variable.
The test fields are defined on spaces of functions vanishing on the Dirichlet part of the domain.
Once the value of the current displacement $\bu^{n+1}$ is obtained from the solution of \eqref{eq:mech}, the Jacobian
$J^{n+1}$ and the deformation gradient $\mathbf{F}^{n+1}$ are computed, and
the nonlinear reaction-diffusion system is solved to find $V^{n+1}, w^{n+1}$:

\begin{equation}\label{eq:elec}
 \begin{aligned}
&\dfrac{1}{\Delta t}  \int_{\Omega_0}   (V^{n+1}- V^n)\xi \; {\rm d} \bX+
 \int_{\Omega_0} \dfrac{1}{J^{n+1}}  (\mathbf{F}^{n+1} )^{-1} \mathbf{D} ( \mathbf{F}^{n+1} )^{- \rm T} \nabla V^{n+1} \cdot \nabla \xi \; {\rm d} \bX -
 \int_{\Omega_0}     I^{n+1} \xi \; {\rm d} \bX- \\
& \int_{\Omega_0}    I_{ \rm app} \xi \; {\rm d} \bX =0  , \quad \forall \xi \in \mathcal{V}_{0,h},\\
&\dfrac{1}{\Delta t}  \int_{\Omega_0}  (w^{n+1}- w^n)\phi \; {\rm d} \bX- \int_{\Omega_0}  H^{n+1} \phi \; {\rm d} \bX =0, \quad \forall \phi \in \mathcal{W}_{0,h},
 \end{aligned}
 \end{equation}
where the functions $I^{n+1}$ and $H^{n+1}$ are given by
 $I^{n+1}=I^{n+1}(V^{n+1}, w^{n+1})$ and $H^{n+1}=H^{n+1}(V^{n+1}, w^{n+1})$ and the new values of $\gamma^{n+1}_l, \gamma^{n+1}_t$ are computed
 using Eq. \eqref{eq:activation} and \eqref{eq:gammat}, respectively.
Both problems \eqref{eq:mech} and \eqref{eq:elec} are nonlinear problems and at each timestep $t^{n+1}$, a nested Newton-Raphson iterative
scheme must be applied to their linearized counterparts to find approxmate solution of the field variables.
Equation \eqref{eq:mech} must be linearized in order to apply the Newton's method.
The Fr\'{e}chet devivatives of each term of Eq.
\eqref{eq:mech} are computed as follows. From the first line of Eq. \eqref{eq:mech}:
\begin{equation*}
\begin{aligned}
 < d ( \int_{\Omega_0} \mu J_a \mathbf{F} ( \mathbf{F}_a )^{-1} ( \mathbf{F}_a )^{- \rm T} : \nabla \bv \; {\rm d} \bX  )_{\bu} , \delta \bu >
 = \int_{\Omega_0} \mu J_a d \mathbf{F}  ( \mathbf{F}_a )^{-1} ( \mathbf{F}_a )^{- \rm T} : \nabla \bv \; {\rm d} \bX ,
 \end{aligned}
\end{equation*}
where $d \mathbf{F}(\bu):=\nabla (d \bu)$.
Recalling that if $A$ is a $2 \times 2$ invertible matrix $A=\left( \begin{array}{cc}
w & x  \\
y & z   \end{array} \right)$, the cofactor matrix is defined as
$\text{Cof}(A)=\left( \begin{array}{cc}
z & -y  \\
-x & w   \end{array} \right)$ so in $2 D$ the identity $A^{-1}=\dfrac{1}{\det(A)} \text{Cof}(A)^{ \rm T}$ holds. Then:
\begin{equation*}
< d ( \int_{\Omega_0} J p \mathbf{F}^{-T} : \nabla \mathbf{v} \; {\rm d} \bX  )_{\bu} , \delta \bu >=
\int_{\Omega_0} p \text{Cof}(d \mathbf{F}): \nabla \mathbf{v} \; {\rm d} \bX,
\end{equation*}
where the realtions $J \mathbf{F}^{- \rm T} = \text{Cof} (\mathbf{F} )$ and $d \text{Cof} (\mathbf{F} ) = \text{Cof} (d\mathbf{F} )$ were used.
Then is computed:
\begin{equation*}
< d ( \int_{\Omega_0} J p \mathbf{F}^{- \rm T} : \nabla \bv \; {\rm d} \bX  )_{p} , \delta p>= \\
\int_{\Omega_0}  J dp \mathbf{F}^{-T} : \nabla \bv \; {\rm d} \bX .
\end{equation*}
Finally, from the identity $\dfrac{\partial J}{\partial \bu}=J \mathbf{F}^{- \rm T} : \dfrac{\partial \mathbf{F}}{\partial \bu}$
it follows:
\begin{equation*}
 < d ( \int_{\Omega_0} (J-1) q \; {\rm d} \bX )_{\bu}, \delta \bu>=\int_{\Omega_0} J \mathbf{F}^{- \rm T} : d \mathbf{F} q \; {\rm d} \bX.
\end{equation*}
Dropping the superscript denoting time discretization,
the linearized equation \eqref{eq:mech} to be solved in the correction $(\delta \bu, \delta p )$ at each  Newton-Raphson iteration is:
Given values at previous time iteration $(\bu^n, p^n)$ and
values at previous Newton-Raphson iteration $(\bu^{n+1}_k, p^{n+1}_k)$ of displacement field and pressure, find the corrections
$(\delta \bu, \delta p)$ such that:
\begin{equation*}\label{eq:linmec}
\begin{aligned}
	&  \int_{\Omega_0} \mu  J^n_a \delta \mathbf{F} ( \mathbf{F}^n_a )^{-1}( \mathbf{F}^n_a )^{- \rm T} : \nabla \bv \; {\rm d} \bX   -
	\int_{\Omega_0} p^{n+1}_k \text{Cof}(\delta \mathbf{F}): \nabla \bv \; {\rm d} \bX -
	\int_{\Omega_0}  J^{n+1}_k \delta p ( \mathbf{F}^{n+1}_k )^{- \rm T} : \nabla \bv \; {\rm d} \bX + \\
	& \int_{\Omega_0} J^{n+1}_k (\mathbf{F}^{n+1}_k ) ^{- \rm T} : \delta \mathbf{F} q \; {\rm d} \bX
	=-R(\bu^{n+1}_k, p^{n+1}_k), \qquad \forall \bv \in \bV_{0,h}, q \in Q_{0,h},
\end{aligned}
\end{equation*}
where $\delta \mathbf{F}=\nabla (\delta \bu)$ and $R(\bu^{n+1}_k, p^{n+1}_k)$ is the residual given by:
\begin{equation*}
\begin{aligned}
R(\bu^{n+1}_k, p^{n+1}_k)= &   \int_{\Omega_0} \mu   J^n_a \mathbf{F}^{n+1}_k ( \mathbf{F}^n_a )^{-1} ( \mathbf{F}^n_a )^{- \rm T} : \nabla \bv \; {\rm d} \bX -
   \int_{\Omega_0}  J^{n+1}_k p^{n+1}_k ( \mathbf{F}^{n+1}_k )^{- \rm T}  : \nabla \bv \; {\rm d} \bX +
 \int_{\Omega_0} (J^{n+1}_k-1)q  \; {\rm d} \bX
\end{aligned}
\end{equation*}
The stopping criterion is:
\begin{equation*}
 \dfrac{\Vert \delta \bu \Vert^2_{\mathbf{H}^1(\Omega_0)}}{\Vert  \bu^{n+1}_k \Vert^2_{\mathbf{H}^1(\Omega_0)}}+
 \dfrac{\Vert \delta p \Vert^2_{ L^2(\Omega_0)}}{\Vert  p^{n+1}_k \Vert^2_{L^2(\Omega_0)}} < \rm{tol_m},
\end{equation*}
where ${ \rm tol_m}$ is a given tolerance and the norm $\Vert \cdot \Vert^2_{\mathbf{H}^1(\Omega_0)}$ is defined as
$\Vert \bu \Vert^2_{\mathbf{H}^1(\Omega_0)}= \Vert \bu \Vert^2_{\mathbf{L}^2(\Omega_0)}+\Vert \nabla \bu \Vert^2_{\mathbf{L}^2(\Omega_0)}$,
while the norm $\Vert \cdot \Vert^2_{\mathbf{L}^2(\Omega_0)}$ is the usual integral squared (vector) norm.

At timestep $t^{n+1}$, after convergence of the mechanical problem,
the new values of displacement and pressure $\bu^{n+1}$ and $p^{n+1}$ are used to compute
 $J^{n+1}$ and $\bF^{n+1}$ entering the reaction-diffusion system \eqref{eq:elec}.
The reaction-diffusion system \eqref{eq:elec} is a nonlinear system in the reaction
electrophysiological functions $I(V,w),H(V,w)$. A nested Newton-Raphson scheme must be used to find an
approximate solution $V^{n+1}, w^{n+1}$ of the linearized
counterpart of \eqref{eq:elec}. At the $k$-th Newton-Raphson iteration,
given the values of the electrophysiological variables
$V^{n}, w^n$ and $V^{n+1,k}, w^{n+1,k}$ respectively at the previous timestep and Netwon-Raphson iteration, it must be solved the following
reaction-diffusion system in the corrections $\delta V, \delta w$ for all test functions $\xi, \phi$ vanishing on the
corresponding Dirichlet part of the domain:
\begin{equation*}
 \begin{aligned}
&\dfrac{1}{\Delta t}  \int_{\Omega_0}   \delta V \xi \; {\rm d} \bX+  \int_{\Omega_0} \dfrac{1}{J^{n+1}}  (\mathbf{F}^{n+1} )^{-1} \mathbf{D} ( \mathbf{F}^{n+1} )^{- \rm T} \nabla \delta V \cdot \nabla \xi \; {\rm d} \bX -
\int_{\Omega_0}    \dfrac{\partial I^{n+1}_k }{\partial V} \delta V \xi \; {\rm d} \bX -\int_{\Omega_0}    \dfrac{\partial I^{n+1}_k }{\partial w} \delta w \xi \; {\rm d} \bX \\
&=-R_V(v^{n+1}_k, w^{n+1}_k),  \quad \forall \xi \in \mathcal{V}_{0,h}, \\
& \dfrac{1}{\Delta t}  \int_{\Omega_0}  \delta w \phi\; {\rm d} \bX- \int_{\Omega_0}  \dfrac{ \partial H^{n+1}_k}{\partial V} \delta V \phi \; {\rm d} \bX-
 \int_{\Omega_0}  \dfrac{ \partial H^{n+1}_k}{\partial w} \delta w \phi \; {\rm d} \bX
=-R_w(V^{n+1}_k, w^{n+1}_k) , \quad \forall \phi \in \mathcal{W}_{0,h},
 \end{aligned}
 \end{equation*}
where the residuals for the two equations are:
\begin{equation*}
\begin{aligned}
 R_v(V^{n+1}_k, w^{n+1}_k)=&\dfrac{1}{\Delta t}  \int_{\Omega_0}  ( V^{n+1,k}-V^n) \xi \; {\rm d} \bX+
 \int_{\Omega_0} \dfrac{1}{J^{n+1}}  (\mathbf{F}^{n+1} )^{-1} \mathbf{D} ( \mathbf{F}^{n+1} )^{- \rm T} \nabla  V^{n+1,k} \cdot \nabla \xi \; {\rm d} \bX- \\
& \int_{\Omega_0}    I^{n+1}_k  V^{n+1,k} \xi \; {\rm d} \bX-\int_{\Omega_0}    I_{\rm app } \xi \; {\rm d} \bX ,
 \end{aligned}
\end{equation*}
and
\begin{equation}
  R_w(V^{n+1}_k, w^{n+1}_k)= \dfrac{1}{\Delta t}  \int_{\Omega_0}  ( w^{n+1,k}-w^n) \phi\; {\rm d} \bX-
 \int_{\Omega_0}   H^{n+1}_k \phi \; {\rm d} \bX .
\end{equation}
The stopping criterion is
$\Vert \delta V \Vert^2_{L^2(\Omega_0)}+ \Vert \delta w \Vert^2_{L^2(\Omega_0)} < \rm{tol_e}$, where $\rm{tol_e}$ is a
given tolerance.





\end{appendix}


\end{document}